\documentclass[acmtog]{acmart}

\usepackage{nicefrac}
\usepackage{hyperref}
\usepackage{multirow}
\usepackage{graphicx}
\usepackage{subfigure}
\usepackage{array}
\usepackage{letltxmacro}
\usepackage{amsmath}
\usepackage{mathtools}
\usepackage{bm}
\usepackage[nolist]{acronym}

\begin{acronym}
\acro{VRR}{variable refresh rate}
\acro{CFF}{Critical Flicker Frequency}
\acro{TLM}{temporal light modulation}
\acro{TCSF}{Temporal Contrast Sensitivity Function}
\acro{CSF}{Contrast Sensitivity Function}
\acro{fMRI}{functional magnetic resonance imaging}
\acro{ICDM}{International Committee Display Metrology}
\acro{IDMS}{Information Display Measurements Standard}
\acro{GPU}{Graphics Processing Unit}
\acro{V-blank}{Vertical Blanking Interval}
\acro{RMSE}{root-mean-square-error}
\end{acronym}

\AtBeginDocument{%
  \providecommand\BibTeX{{%
    \normalfont B\kern-0.5em{\scshape i\kern-0.25em b}\kern-0.8em\TeX}}}

\copyrightyear{2025}
\acmYear{2025}
\setcopyright{rightsretained}
\acmConference[SIGGRAPH Asia'25]{SIGGRAPH Asia 2025 Conference Papers}{December 15--18, 2025}{Hong Kong}
\acmBooktitle{SIGGRAPH Asia 2025 Conference Papers (SA Conference Papers
'25), December 15--18, 2025, Hong Kong, China}
\acmDOI{10.1145/3757377.3763825}
  
\acmSubmissionID{1077}

\begin{document}
\definecolor{yancheng}{RGB}{0,100,0}
\newcommand{\figref}[1]{Figure~\ref{fig:#1}}
\newcommand{\secref}[1]{Section~\ref{sec:#1}}
\newcommand{\algoref}[1]{Algorithm~\ref{algo:#1}}
\newcommand{\chapterref}[1]{Chapter~\ref{chapter:#1}}
\newcommand{\appref}[1]{Appendix~\ref{app:#1}}
\newcommand{\tableref}[1]{Table~\ref{tab:#1}}
\newcommand{\etal}{et al.\xspace}
\newcommand{\degpers}{\,\nicefrac{deg}{s}\xspace}
\newcommand{\degperssq}{\,\nicefrac{deg}{s\textsuperscript{2}}\xspace}
\newcommand{\degree}{$^{\circ}$\xspace}
\newcommand{\Hz}{\,Hz\xspace}
\newcommand{\csdm}{\,cd/m$^2$}
\newcommand{\fourier}{\mathfrak{F}}
\newcommand{\infourier}{^{\mathfrak{F}}}

\LetLtxMacro{\originaleqref}{\eqref}
\renewcommand{\eqref}[1]{Eq.~\originaleqref{eq:#1}}

\newcommand{\todo}[1]{\textcolor{red}{\textbf{todo: #1}}}
\newcommand{\RM}[1]{\textcolor{brown}{\textnormal{(Rafal) #1}}}
\newcommand{\AB}[1]{\textcolor{purple}{\textnormal{(Ali) #1}}}
\newcommand{\MA}[1]{\textcolor{blue}{\textnormal{(Maliha) #1}}}
\newcommand{\MAtext}[1]{\textcolor{blue}{\textnormal{#1}}}

\newcommand{\edit}[1]{\textcolor{black}{#1}}

\definecolor{yancheng}{RGB}{0,0,0}

\newcommand{\ourmethod}{elaTCSF}

\newcommand{\code}[1]{\texttt{#1}}

\newcommand{\ind}[1]{\text{#1}}

\newcommand{\rawim}{\mathbf{I}}
\newcommand{\rawimic}{\mathbf{I}_{i,c}}
\newcommand{\dark}{d}
\newcommand{\queff}{k}
\newcommand{\scrad}{\mathbf{\Psi}}
\newcommand{\readnoise}{\sigma^2_{\mathrm{read}}}
\newcommand{\adcnoise}{\sigma^2_{\mathrm{adc}}}
\newcommand{\readnoisec}{\sigma^2_{\mathrm{read,c}}}
\newcommand{\adcnoisec}{\sigma^2_{\mathrm{adc,c}}}
\newcommand{\rerad}{\mathbf{X}}
\newcommand{\reradi}{\mathbf{X}_i}
\newcommand{\reradic}{\mathbf{X}_{i,c}}
\newcommand{\noMTFrad}{\mathbf{L}^{M}}
\newcommand{\noMTFradc}{\mathbf{L}^{M}_c}
\newcommand{\noMTFradcp}{\mathbf{L}'^{M}_c}
\newcommand{\noMTFradcestimate}{\hat{\mathbf{L}}^{M}_c}
\newcommand{\noMTFradcpestimate}{\hat{\mathbf{L}}'^{M}_c}
\newcommand{\psf}{P}
\newcommand{\freq}{\boldsymbol{\omega}}
\newcommand{\vig}{\mathbf{V}}
\newcommand{\VCradc}{\mathbf{L}^{V}}
\newcommand{\VCradcestimatenoc}{\hat{\mathbf{L}}^{V}}
\newcommand{\VCradcestimate}{\hat{\mathbf{L}}^{V}_c}
\newcommand{\GHD}{\mathbf{L}^{D}}
\newcommand{\GHDroot}{\mathcal{L}^{D}}
\newcommand{\GHestimateD}{\hat{\mathbf{L}}^{D}}
\newcommand{\GHestimateDroot}{\hat{\mathcal{L}}^{D}}
\newcommand{\GHestimate}{\hat{\mathbf{L}}^{D}_c}
\newcommand{\homo}{\mathbf{H}}
\newcommand{\ccmatrix}{\mathbf{M}}
\newcommand{\ccmatrixroot}{\mathcal{M}}
\newcommand{\ccresult}{\mathbf{Y}}
\newcommand{\ccestimate}{\hat{\mathbf{Y}}}
\newcommand{\dsize}{a_\mathrm{d}}
\newcommand{\dcontrast}{c_\mathrm{d}}
\newcommand{\dissizew}{w_\mathrm{d}}
\newcommand{\dissizeh}{h_\mathrm{d}}
\newcommand{\meanmap}{\mathbf{M}_\mathrm{p}}
\newcommand{\backmap}{\mathbf{M}_\mathrm{b}}
\newcommand{\dthreshold}{D_{\mathrm{thr}}}
\newcommand{\nfp}{N_{\text{fp}}}
\newcommand{\dissubp}{\mathsf{k}}

\newcommand{\pix}{\textbf{p}}
\newcommand{\pixscreen}{\textbf{s}}
\newcommand{\pixundistort}{\textbf{g}}
\newcommand{\pixdefect}{\textbf{q}}
\newcommand{\noise}{\eta}
\newcommand{\spfreq}{\textbf{\omega}}


\title{CameraVDP: Perceptual Display Assessment with Uncertainty Estimation via Camera and Visual Difference Prediction}

\author[]{Yancheng Cai}
\email{yc613@cam.ac.uk}
\affiliation{
  \institution{University of Cambridge}
  \streetaddress{William Gates Building, 15 JJ Thomson Avenue}
  \city{Cambridge}
  \postcode{CB3 0FD}
  \country{United Kingdom}
}

\author[]{Robert Wanat}
\email{robwanat@gmail.com}
\affiliation{
  \institution{LG Electronics North America}
  \city{Santa Clara}
  \country{United States of America}
}

\author[]{Rafał K. Mantiuk}
\email{rafal.mantiuk@cl.cam.ac.uk}
\affiliation{
  \institution{University of Cambridge}
  \streetaddress{William Gates Building, 15 JJ Thomson Avenue}
  \city{Cambridge}
  \postcode{CB3 0FD}
  \country{United Kingdom}
}


\begin{abstract}
Accurate measurement of images produced by electronic displays is critical for the evaluation of both traditional and computational displays. Traditional display measurement methods based on sparse radiometric sampling and fitting a model are inadequate for capturing spatially varying display artifacts, as they fail to capture high-frequency and pixel-level distortions. While cameras offer sufficient spatial resolution, they introduce optical, sampling, and photometric distortions. Furthermore, the physical measurement must be combined with a model of a visual system to assess whether the distortions are going to be visible. To enable perceptual assessment of displays, we propose a combination of a camera-based reconstruction pipeline with a visual difference predictor, which account for both the inaccuracy of camera measurements and visual difference prediction. The reconstruction pipeline combines HDR image stacking, MTF inversion, vignetting correction, geometric undistortion, homography transformation, and color correction, enabling cameras to function as precise display measurement instruments. By incorporating a Visual Difference Predictor (VDP), our system models the visibility of various stimuli under different viewing conditions for the human visual system. We validate the proposed CameraVDP framework through three applications: defective pixel detection, color fringing awareness, and display non-uniformity evaluation. Our uncertainty analysis framework enables the estimation of the theoretical upper bound for defect pixel detection performance and provides confidence intervals for VDP quality scores. Our code is available on \textcolor{cyan}{\url{https://github.com/gfxdisp/CameraVDP}}.
\end{abstract}





\begin{teaserfigure}
 \centering
      \includegraphics[width=0.95\linewidth]{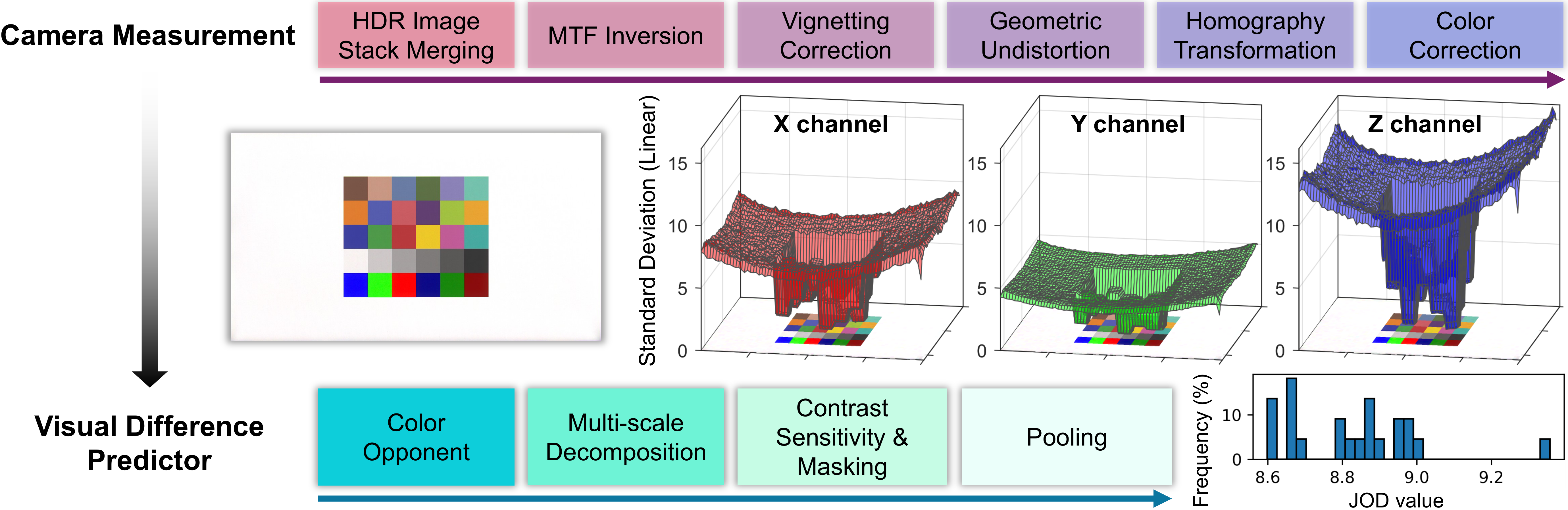}
  \caption{CameraVDP pipeline. The camera first captures distortion-free, color-corrected measurements along with their uncertainty covariance (standard deviations in XYZ space shown). Test and reference measurements are then input to the visual difference predictor to compute a quality score distribution. Values and uncertainties are jointly propagated at each step. (JOD: just-objectionable-differences, see~\cite{mantiuk2021fovvideovdp, mantiuk2024colorvideovdp})}
  \label{fig:teaser}
\end{teaserfigure}
\maketitle

\section{Introduction}
\label{sec:introduction}

Accurate display evaluation is critical for the design and maintenance of display systems, ensuring reliable visual content reproduction across diverse usage scenarios while avoiding visible artifacts. This entails three key requirements: precise display measurement, perceptual modeling incorporating human visual characteristics, and uncertainty estimation in both measurement and modeling.

Display measurement must accurately capture patterns of high or low spatial frequencies to support both pixel-level evaluation (e.g., defective pixels) and full-screen assessment (e.g., uniformity). The ideal solution is to use high-resolution 2D photometers or colorimeters; however, such devices are not affordable for small research laboratories, and their processing stack is proprietary. High-resolution photographic (prosumer) cameras that capture RAW images offer a practical alternative, but they face several challenges: (1) with only three color filters, they cannot reconstruct the input optical power spectrum; (2) camera lenses introduce geometric distortions; (3) their dynamic range is limited; (4) vignetting introduces non-uniformity in captured images; (5) and high spatial frequencies are attenuated or lost due to optical aberrations. To overcome these limitations, we propose an open-source measurement pipeline for mirrorless cameras, customized for the measurement of displays. It takes advantage of regular display pixel layout and a limited set of subpixel primaries. The pipeline integrates HDR image stack merging, MTF inversion, vignetting correction, geometric undistortion, homography transformation, and color correction. 

A display defect or artifact matters only when it is visible to the human eye. The visibility of such artifacts can be predicted by visual difference predictors \cite{daly1992visible,mantiuk2024colorvideovdp}. Those, however, require perfectly aligned and physically calibrated test and reference images. The goal of our camera correction pipeline is to provide such images for visual difference predictors. \edit{We employ ColorVideoVDP \cite{mantiuk2024colorvideovdp}, which models contrast masking and contrast sensitivity in a range of luminance from 0.001 to 10000\csdm{}, making it capable of handling both SDR and HDR displays.}



A measurement instrument needs to provide an estimate of the measurement error it makes in order to ensure it is accurate enough for a given task. We include this capability in our method by analytically modeling the propagation of sensor noise (uncertainty) through all the stages of camera image correction. Furthermore, we train multiple-versions of VDP (ColorVideoVDP, \cite{mantiuk2024colorvideovdp}) on the XR-DAVID dataset to estimate its prediction error via Monte Carlo simulation. This allows us to use the CameraVDP to robustly detect defective pixels, estimate the visibility of color fringing artifacts and display non-uniformity. 



In summary, our contributions are as follows:
\begin{itemize}
\item An \textcolor{yancheng}{open source} pipeline for correcting distortions introduced by mirrorless cameras, which takes advantage of display-specific priors. Such a pipeline can be combined with a VDP to estimate the visibility of the distortions. 
\item A forward uncertainty propagation framework, incorporating analytical normal distribution-based estimation for camera measurement uncertainty and Monte Carlo simulation for VDP uncertainty.
\item We demonstrate CameraVDP's accuracy and the importance of uncertainty estimation through three applications: defect pixel detection, color fringing, and uniformity assessment.
\end{itemize}

\section{Related Work}
\subsubsection*{Geometric and radiometric camera calibration}
Geometric and radiometric camera calibration integrate models of optics, sensors, and image processing to bridge the gap between raw sensor measurements and physically accurate image representations. HDR imaging~\cite{debevec2023recovering, hanji2020noise} reconstructs relative scene radiance from multi-exposure sequences, preserving details in both shadows and highlights. Color correction~\cite{finlayson2015color} maps RGB values captured with camera primaries into trichromatic color values for a given color matching functions (e.g., CIE XYZ 1931). Geometric distortion correction~\cite{tang2017precision, weng1992camera} relies on parametric models to rectify lens distortions. The camera spatial frequency response (SFR/MTF)~\cite{burns2022updated} (ISO 12233 standard), typically estimated using the slanted-edge method, quantifies the spatial resolution of imaging systems and serves as a key metric for optical performance evaluation. This work integrates these calibration techniques into a unified pipeline, enabling commercial cameras to accurately capture high-fidelity content for HDR displays, supporting downstream tasks such as display uniformity and color fringing assessment. We rely on an explainable analytical model rather than a differential network \cite{Tseng_2021}, as the former is guaranteed to generalize to any content and is more appropriate for accurate camera measurements.
\edit{
\subsubsection*{Display measurement standard}
The current display measurement standard, IDMS v1.3 \cite{SID_2025}, specifies procedures for measuring different aspects of displays, such as uniformity, contrast, color gamut, and others. The standard describes the use of imaging luminance measurement devices (ILMDs). Those are typically proprietary instruments, calibrated in photometric or colorimetric units, which serve a similar purpose as the ``camera'' part of our CameraVDP. However, besides their high cost, those instruments often come with limitations: limited dynamic range, field of view, and no methods to precisely align captured information with display pixels. Our camera correction pipeline (\secref{measurement}) will address those shortcomings. Although the standard mentions limited use of visual models (e.g., using TCSF for flicker visibility, sec.~10.6), it does not explain how ILMD/camera measurements can be used with a spatial model of human vision (e.g., VDP).}

\subsubsection*{Modeling visible differences}
Due to the high cost of subjective human (expert) evaluation, objective computational models of human visual perception are essential. Accurate models typically incorporate prior knowledge of the human visual system (HVS)~\cite{schade1956optical} and psychophysical data, enabling better generalization to unseen scenarios. The contrast sensitivity functions (CSFs) characterizes the HVS sensitivity—defined as the inverse of contrast detection threshold—to variations in color~\cite{ashraf2024castlecsf}, luminance~\cite{mustonen1993effects}, area~\cite{rovamo1993modelling}, temporal frequency~\cite{cai2024elatcsf, watson1986temporal}, and spatial frequency~\cite{barten2003formula}. Based on CSF and contrast masking modeling, visual difference predictors (VDPs) such as DCTune~\cite{watson1993dctune}, VDP~\cite{daly1992visible}, ColorVideoVDP~\cite{mantiuk2024colorvideovdp}, and HDR-VDP-3~\cite{mantiuk2023hdr} have been developed to address more complex image and video content. Existing VDPs require perfectly aligned test and reference images, both calibrated in physical units. Our camera correction pipeline provides such alignment and calibration, making camera-capture images a suitable input to VDPs.


\subsubsection*{Uncertainty quantification}
\edit{Uncertainty is typically categorized into two types: aleatoric (irreducible, e.g., uncertainty of VDP) and epistemic (reducible, e.g., camera noise)~\cite{der2009aleatory}}. Its quantification generally falls into two approaches: forward propagation and inverse assessment. Forward propagation estimates overall system uncertainty based on input variability, using methods such as Monte Carlo simulations~\cite{kroese2013handbook} and surrogate models~\cite{ranftl2021bayesian}. Inverse assessment corrects model bias or calibrates parameters using observed data, exemplified by modular~\cite{kennedy2001bayesian} and fully Bayesian~\cite{bayarri2009modularization} approaches. In computational imaging, common approaches to uncertainty quantification include Markov Chain Monte Carlo (MCMC) sampling~\cite{bardsley2012mcmc, broderick2020hybrid}, Bayesian hypothesis testing~\cite{repetti2019scalable}, variational inference~\cite{sun2021deep,arras2019unified,blei2017variational,rezende2015variational,ekmekci2021does}, and Bayesian neural networks~\cite{ongie2020deep,xue2019reliable}.

However, most of these methods target deep learning–based reconstruction tasks. While~\citeN{hagemann2022inferring} addresses bias and uncertainty in camera calibration via a resampling-based estimator, a unified framework for uncertainty propagation through HDR merging, geometric undistortion, homography transformation, and color calibration remains absent. Moreover, uncertainty modeling is entirely unexplored in perception modeling~\cite{cai2025computer}.

\begin{figure}[t]
  \centering
      \includegraphics[width=\linewidth]{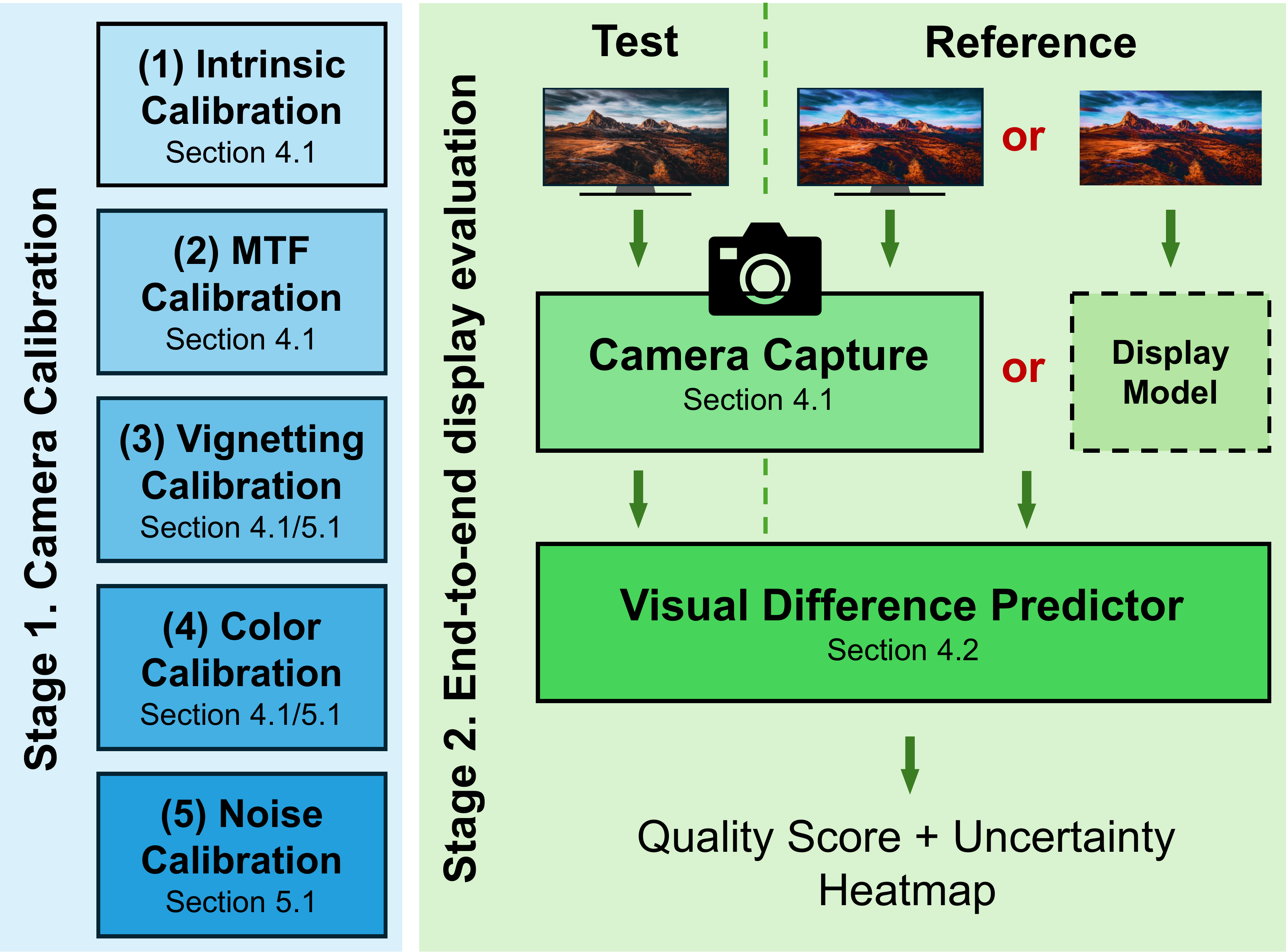}
  \caption{\textcolor{yancheng}{CameraVDP as an end-to-end display evaluation system.}}
  \label{fig:cameravdp_system}
\end{figure}

\begin{color}{yancheng}
\section{Overview of CameraVDP}
CameraVDP is a complete camera-based display assessment system, as shown in \figref{cameravdp_system}, consisting of a camera calibration stage and an end-to-end evaluation stage. Besides the test display, the required equipment includes a digital camera with pixel-shift mode (or a monochrome camera with color filters to avoid demosaicing), and a spectroradiometer or colorimeter.

The camera calibration stage is done in five steps: (1) calibration of intrinsic and lens distortion parameters; (2) MTF (SFR) measurement; (3) calibration of the vignetting map; (4) estimation of the color correction matrix; (5) estimation of noise parameters. Steps (1)–(3) must be repeated for each camera–lens–focal length configuration, while (4) requires recalibration for each camera–lens–display combination. The display needs to be warmed up (30\,minutes) before the procedures in steps (1), (3) and (4). 
Camera measurement distance should be recorded, as focal length variations may cause a slight change of intrinsic camera parameters. The calibration stage is typically completed within two hours.

The evaluation stage (see the right portion of \figref{cameravdp_system}) consists of camera capture and a Visual Difference Predictor (VDP). The test is the image shown on the display under evaluation, directly captured by the camera. The reference can either be (i) a higher-quality display captured by the same camera, or (ii) an image with correct absolute colorimetric values (e.g., CIE XYZ), where the values are obtained using an idealized display model (e.g., BT.2100 for HDR content). After applying camera corrections (Section~\ref{sec:measurement}), the VDP  (Section~\ref{sec:visual-difference-predictor}) produces a distribution of quality scores with uncertainty estimates, and a visual difference heatmap. During capture, OpenCV ArUco markers are first displayed to estimate extrinsic parameters, followed by the test image; this procedure typically can be completed within 10 minutes.

\end{color}

\section{Methodology}
\label{sec:methodology}
\begin{figure*}[t]
 \centering
      \includegraphics[width=\linewidth]{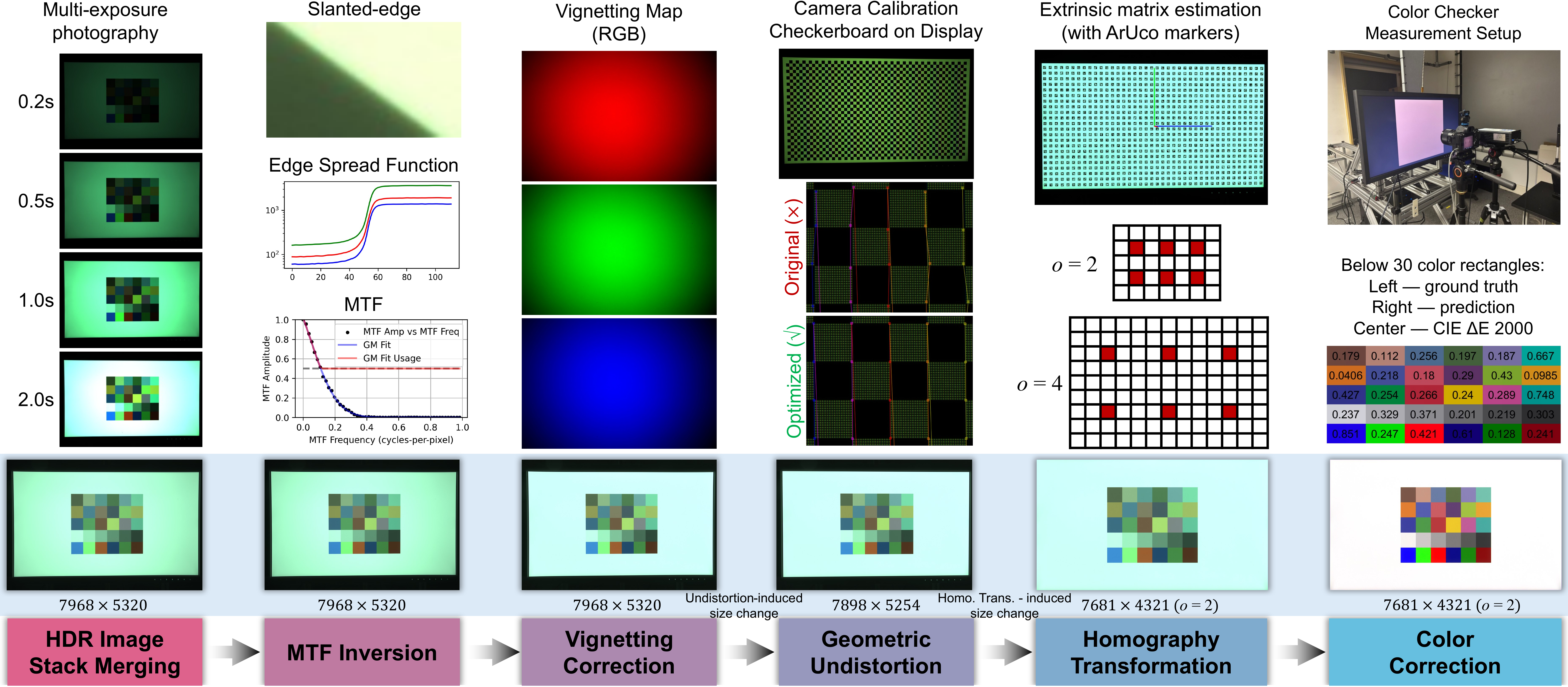}
  \caption{The complete camera measurement pipeline is shown from left to right, with each column representing a key step and the bottom row illustrating transformation examples. Column 1: HDR acquisition via merging multi-exposure images (4 exposure levels). Column 2: MTF inversion to correct lens aberrations. Column 3: Vignetting correction, exemplified by the vignetting map of the Sony $\alpha$7R III with FE \textcolor{yancheng}{1.8/35mm} lens. Column 4: Intrinsic and distortion parameter estimation using full-screen checkerboard patterns for geometric undistortion (crop empty edges caused by undistortion).  Column 5: Homography estimation of extrinsic parameters using full-screen OpenCV ArUco markers. 
  \textcolor{yancheng}{The display pixel oversampling factor $o$ denotes the distance, in image pixels, between the centers of original display pixels (red squares) in the oversampled image after the homography transformation.}
  Examples with \textcolor{yancheng}{$o=2,4$} are shown, where red squares indicate display pixel centers. Column 6: Color correction by mapping camera RGB to measured XYZ.}
  \label{fig:pipeline}
\end{figure*}

Our approach combines camera-based measurements with a visual difference predictor (VDP). The goal of the camera measurements is to achieve sufficient accuracy by correcting the inherent distortions of standard cameras. The complete workflow, shown in~\figref{pipeline}, comprises six key steps. Based on these measurements, the VDP incorporates a model of human vision to convert physical differences into perceived visual differences. As with any physical measurement, including those obtained using camera sensors, measurement uncertainty is inevitable. Additionally, the VDP introduces prediction uncertainty due to its dependence on the training data. A combined estimation of these uncertainties is essential to assess the reliability of the predicted values.

\subsection{Camera correction}
\label{sec:measurement}
Our camera measurement correction pipeline is designed for cameras equipped with pixel-shift (sensor-shift) feature, which increases resolution and reduces noise and demosaicing artifacts. For that reason, we do not model demosaicing in our pipeline. We also assume that the measured display has three primaries (i.e., RGB subpixels). 

Uncertainty estimation is based on three assumptions: the noise is independent across (1) exposures, (2) individual pixels, and (3) RGGB camera subpixels. 

Our derivation involves extensive notation, with full definitions provided in Table S1 (supplementary materials).


\paragraph{\textbf{HDR Image Stack Merging and Noise Model}}
The dynamic range of HDR displays often exceeds that of camera sensors, necessitating the capture of multiple exposures. Merging exposures can also reduce noise (via averaging). The RAW (digital) sensor values $\rawim$ for a pixel $\pix\in\mathbb{R}^2$ can be modeled as \cite{aguerrebere2014best}:
\begin{equation}
    \rawim_c(\pix) \sim \queff_c \{\mathrm{Pois} ((\scrad_c(\pix)+\dark)t)g + \mathcal{N} (0,\readnoisec)g+\mathcal{N} (0,\adcnoisec)\}
    \label{eq:camera-noise},
\end{equation}
where $c\in\{\mathsf{r},\mathsf{g},\mathsf{b}\}$ is the color channel index, $t$ is the exposure time, $g = \mathrm{ISO}/100$ is sensor's gain, $d$ is the dark current noise, and  $\scrad$ is the scene radiance. $\queff_c$, $\readnoisec$ and $\adcnoisec$ are the noise parameters, which we estimate in~\secref{camera-noise-param-est}. We measured the dark noise $\dark$ for our test camera by capturing $I$ with the lens blocked ($\scrad \approx 0$) at room temperature. As the dark noise remained below 0.1\% within our exposure range, it was omitted from further analysis.

Before merging a stack of RAW exposures $i = 1...N$, we need to compensate for differences in exposure time and gain. The relative radiance $\rerad_c(\pix)$ is defined as $\rerad_c(\pix) \coloneqq \frac{\rawim_c(\pix)}{tg}$. We can formulate the reconstruction of the scene radiance by incorporating the noise model from \eqref{camera-noise} and solving for the maximum likelihood estimation (MLE)~\cite{aguerrebere2014best}. However, solving for MLE is impractical for large sets of high-resolution images. As noted by \citeANP{hanji2020noise} \citeNN{hanji2020noise}, photon noise dominates in modern cameras, especially at long exposure times used for measurements. Therefore, we can assume a simplified noise model $\rerad_{c}(\pix) \sim \queff_c \mathrm{Pois}(\scrad_c(\pix)t)/t$. \textcolor{yancheng}{A closed-form estimator for this model is given by \cite{hanji2020noise}:}
\begin{equation}
    \overline{\scrad}_c(\pix) = \frac{\sum_{i=1}^{N} \queff_c x_{i,c}(\pix) t_{i}}{\sum_{i=1}^{N} t_{i}}
    \label{eq:hanji_estimator},
\end{equation}
where $x_{i,c}$ denotes a single measurement ($i = 1...N$) of the random variable $\rerad_{c}$. 

Given the sufficient photon count received by the camera during exposure under typical display luminance, the Poisson distribution can be approximated by a normal distribution. Then, the  relative radiance $\rerad_{c}(\pix)$ follows a normal distribution: 
\begin{equation}
\small
    \rerad_c(\pix) \sim \mathcal{N} \left(\queff_c \scrad_c(\pix),\frac{\queff^2_c\scrad_c(\pix)}{t}+\frac{\queff^2_c\readnoisec}{t^2}+\frac{\queff^2_c\adcnoisec}{t^2g^2}\right)
    \label{eq:relative_radiance_distribution}.
\end{equation}
The estimated distribution of radiance $\hat{\scrad}$ from exposure stack is:
\begin{equation}
\small
    \hat{\scrad}_c(\pix)\sim \mathcal{N} \left(\frac{\sum_{i=1}^{N} \queff_c x_{i,c}(\pix) t_{i}}{\sum_{i=1}^{N} t_{i}},\frac{\sum_{i=1}^N [\queff^2_cx_{i,c}(\pix)t_i+\queff_c^2\sigma^2_{\mathrm{read},c}+\queff_c^2\sigma^2_{\mathrm{adc},c}/g_i^2]}{(\sum_{i=1}^{N} t_{i})^2}\right)
    \label{eq:scene_radiance_distribution}.
\end{equation}

\begin{figure}[H]
  \centering
      \includegraphics[width=\linewidth]{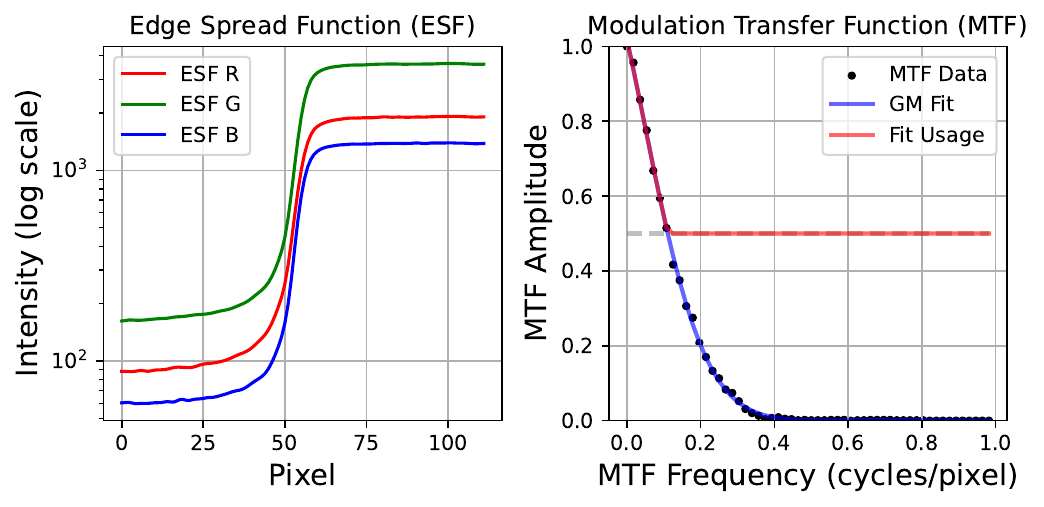}
  \caption{Edge Spread Function (ESF) and Modulation Transfer Function (MTF) results. To prevent excessive amplification of noise, the MTF is constrained to a minimum value of 0.5 (indicated by the red line). Images were captured of the diagonal black-white edge at the center of a Siemens star rotated by \(45^\circ\). 
}
  \label{fig:ESF_MTF_results}
\end{figure}
\paragraph{\textbf{MTF Inversion}} 
Due to lens aberrations, imperfect focus and lens glare, imaging systems cannot accurately reproduce high spatial frequency details, resulting in blur. The Modulation Transfer Function (MTF), also known as the Spatial Frequency Response (SFR), characterizes an imaging system’s ability to resolve fine spatial details. Classical measurement methods include edge-based (e-SFR)~\cite{kerr2024recommendations} and sinewave-based analysis (s-SFR)~\cite{loebich2007digital}.

The scene radiance $\scrad_c(\pix)$ affected by blur and glare can be modeled as the convolution of the clean scene radiance $\noMTFradc(\pix)$ and the camera point spread function (PSF) $\psf(\pix)$, plus noise $\eta(\pix)$:
\begin{equation}
    \scrad_c (\pix) = (\noMTFradc * \psf)(\pix) + \eta(\pix)
    \label{eq:MTF_convolution},
\end{equation}
where $*$ denotes convolution. The MTF $M(\freq)$ is the modulus of the Fourier transform of the PSF $\psf(\pix)$~\cite{burns2022updated}, $\freq\in\mathbb{R}^2$ is spatial frequency in cycles-per-pixel. Taking the Fourier transform $\mathcal{F}$ on both sides of~\eqref{MTF_convolution} yields:
\begin{equation}
    \mathcal{F} (\scrad_c) = \mathcal{F} (\noMTFradc) M(\freq) + \eta'(\freq)
    \label{eq:fourier_MTF},
\end{equation}
Considering the noise, we use Wiener deconvolution to obtain the deglared and deblurred estimate of $\noMTFradc$:
\begin{equation}
\noMTFradcestimate(\pix) = \mathcal{F}^{-1}\left(\mathcal{F}(\hat{\scrad}_c) G_c(\freq)\right)(\pix)
    \label{eq:wiener_deconvolution}\,,
\end{equation}
where the Wiener filter is ($^{*}$ is the conjugate operator):
\begin{equation}
G_c(\freq)=\frac{M^{*}(\freq) S_{\hat{\scrad}_c}(\freq)}{|M(\freq)|^{2} S_{\hat{\scrad}_c}(\freq)+N_{\hat{\scrad}_c}(\freq)}
    \label{eq:wiener_filter}\,.
\end{equation}
$S_{\hat{\scrad}_c}(\freq)$ and $N_{\hat{\scrad}_c}(\freq)$ represent the power spectral densities (PSD) of the signal and noise, see supplementary materials for details.

The MTF was assumed to be isotropic. We measured $M(\freq)$ using the slanted-edge method (\figref{pipeline}, column 2), and fitted using: 
\begin{equation}
\small
M'(\freq)=a_{1} \exp \left(-\left(\frac{\freq-b_{1}}{c_{1}}\right)^{2}\right)+a_{2} \exp \left(-\left(\frac{\freq-b_{2})}{c_{2}}\right)^{2}\right)\,, 
\label{eq:MTF_fitting}
\end{equation}
where $a_1, a_2, b_1, b_2, c_1, c_2$ are fitting parameters. To suppress noise amplification, \edit{we restrict the minimum modulation to be 0.5 ($M(\freq) = \max(M'(\freq), 0.5)$), as shown in~\figref{ESF_MTF_results}. The minimum modulation also makes the effects of anisotropy and spatial variance of MTF negligible.}

Since the Fourier transform is a linear operation, $\hat{L}_{c}(\pix)$ follows a normal distribution. The mean of $\noMTFradcestimate(\pix)$ is given by~\eqref{wiener_deconvolution}:
\begin{equation}
    \mu_{\noMTFradcestimate}(\pix) = \mathcal{F}^{-1}\left(\mathcal{F}(\mu_{\scrad_c})) G(\freq)\right)(\pix)
    \label{eq:mu_MTF}\,,
\end{equation}
Under the assumption of spatial white noise, the variance is:
\begin{equation}
\sigma^{2}_{\noMTFradcestimate}(\pix) \approx \sigma^{2}_{\scrad_c}(\pix) \iint\left|G\left(\freq\right)\right|^{2} d\freq
\label{eq:variance_MTF}.
\end{equation}
The derivation can be found in Section S1 of the supplementary.

\paragraph{\textbf{Vignetting Correction}} Vignetting refers to the reduction in image brightness toward the edges compared to the center. It can be categorized into mechanical, optical, natural and pixel vignetting, and is associated with factors such as partial occlusion at the lens periphery and the cosine fourth-power falloff~\cite{asada1996photometric}. 

Although many reference-free vignetting correction methods~\cite{zheng2008single,kang2000can} exist, flat-field correction remains the most widely used approach. It involves capturing the image $\rawim$ of a uniformly illuminated and color-consistent light source $\rawim_{\mathrm{flat}}$:
\begin{equation}
    \rawim_c(\pix) = \vig_c(\pix)\rawim_{\mathrm{flat},c} + \epsilon_c(\pix),
    \label{eq:vignetting_source}
\end{equation}
where $\vig_c(\pix) \in (0,1]$ is the vignetting function and $\epsilon_c(\pix)$ is noise. To compensate for the impact of dust in the optical system, the vignetting function is estimated without smoothing:
\begin{equation}
    \hat{\vig}_c(\pix) = \noMTFradcestimate(\pix) / \mathrm{max} (\noMTFradcestimate(\pix)).
    \label{eq:vignetting_map}
\end{equation}
The vignetting correction result is:
\begin{equation}
    \VCradcestimate(\pix)\sim \mathcal{N} \left(\frac{\mu(\hat{L}_c(\pix))}{\hat{\vig}_c(\pix)}, \frac{\sigma^2(\hat{L}_c(\pix))}{{\hat{\vig}_c}^2(\pix)}\right)
    \label{eq:vignetting_correction}.
\end{equation}

\paragraph{\textbf{Geometric Undistortion and Homography Transformation}}
We aim to find a mapping $m: \pixscreen \to \pix$, from screen pixel coordinates $\pixscreen$ to camera pixel coordinates  $\pix$, such that:
\begin{equation}
    \GHestimate(\pixscreen) = \sum_{\delta{\in}\Omega} \VCradcestimate(m(\pixscreen)+\delta)R(\delta)
    \label{eq:remapping}.
\end{equation}
where $R(\delta)$ is the resampling kernel and $\Omega$ is the neighborhood of a pixel. $m$ is composed of homography $h: \pixscreen \to \pixundistort$ and geometric undistortion $u: \pixundistort \to \pix$, where $\pixundistort$ is the undistorted coordinate.

Homography transformation requires displaying multiple OpenCV ArUco markers~\cite{garrido2014automatic} on screen (see \figref{pipeline}, column 5) to obtain the homography matrix $\homo\in\mathbb{R}^{3 \times 3}$. Then the screen coordinate can be mapped to an undistorted coordinate with:
\begin{equation}
    \mathbf{\pixundistort}^{h}=\mathbf{H} \mathbf{\pixscreen}^{h}
    \label{eq:remapping_homography},
\end{equation}
where $\mathbf{\pixundistort}^{h}, \mathbf{\pixscreen}^{h}$ are the homogeneous coordinates of $\mathbf{\pixundistort}$ and $\mathbf{\pixscreen}$.

To support applications requiring subpixel color structures (e.g., chromatic aberration), display pixels are supersampled with a factor of $o$ ($\geq 3$), as shown in the fifth column of \figref{pipeline}. 

For geometric undistortion, we adopt the Brown-Conrady distortion model, comprising radial and tangential distortions:
\begin{equation}
\small
u(\pixundistort) = \pixundistort \cdot \left( 1 + k_1 \|\pixundistort\|^2 + k_2 \|\pixundistort\|^4 + k_3 \|\pixundistort\|^6 \right)
+ \begin{bmatrix}
2 p_1 \pixundistort_x \pixundistort_y + p_2 \left( \|\pixundistort\|^2 + 2 \pixundistort_x^2 \right) \\
p_1 \left( \|\pixundistort\|^2 + 2 \pixundistort_y^2 \right) + 2 p_2 \pixundistort_x \pixundistort_y
\end{bmatrix}
\label{eq:remapping_undistortion},
\end{equation}
where $k_1,k_2,k_3$ are the radial distortion parameters, $p_1, p_2$ are the tangential distortion parameters, $\|\pixundistort\|^2 = \pixundistort_x^2 + \pixundistort_y^2$.

Using OpenCV for camera calibration on the test display showing a checkerboard pattern, we obtained the intrinsic camera matrix and distortion parameters. Note that the origin of $\pixscreen$ is at the display center, with the $x$-axis pointing right and the $y$-axis pointing upward, using pixel pitch as the unit.

Additionally, we found that when multiple camera pixels capture a single display pixel (with oversampling), OpenCV fails to accurately detect the corner points. To address this, we first identify all display pixel coordinates via local maxima. Then, the coordinates of the detected checkerboard corners are updated to the average of the coordinates of the two white display pixels (not in the same checkerboard block) closest to the original detected corner position. (see~\figref{pipeline}, column 4). When calibrating for the Sony FE \textcolor{yancheng}{2.8/90mm} lens on the Eizo display, our improved method reduced the reprojection error from 0.315 to 0.018, achieving an 18-fold improvement compared to the OpenCV calibration (with subpixel refinement).

When estimating uncertainty, the resampling operation introduces local averaging that compromises the independence of neighboring pixels. To simplify subsequent analysis, we conservatively estimate the variance by resampling it in the same manner as the mean (\eqref{remapping}). Although this may slightly overestimate variance in some regions, it preserves the pixel independence assumption, facilitating Monte Carlo sampling in downstream visual difference predictors.

\paragraph{\textbf{Color Correction}} Cameras cannot capture colors identical to those perceived by the human eye due to differences in spectral sensitivity between the eye and the camera sensor, making color correction essential~\cite{finlayson2015color}. The radiance $\GHestimate$ recorded by the camera for a single display pixel is:
\begin{equation}
\GHestimate = \sum _\dissubp P_\dissubp\int _{\lambda}E_\dissubp(\lambda)C_c(\lambda)d\lambda 
    \label{eq:camera_color_rgb},
\end{equation}
where $\lambda$ is the wavelength, $E_\dissubp(\lambda)$ is the spectral power distribution (SPD) of \textcolor{yancheng}{display} subpixel $\dissubp \in \{\mathsf{R}, \mathsf{G}, \mathsf{B}\}$ at unit intensity, $P_\dissubp$ is the linearized \textcolor{yancheng}{display} subpixel value, and $C_c(\lambda)$ denotes the spectral sensitivity of camera channel $c \in \{\mathsf{r}, \mathsf{g}, \mathsf{b}\}$. Similarly, the trichromatic CIE XYZ value (related to cone responses) can be expressed as:
\begin{equation}
    \ccresult  = \sum _\dissubp P_\dissubp \int_{\lambda }E_\dissubp(\lambda)\bar{\mathbf{S} } (\lambda )d\lambda
    \label{eq:color_xyz}, 
\end{equation}
where $\ccresult=[X,Y,Z]'$ and $\bar{\mathbf{S} } (\lambda ) = [\bar{x}(\lambda ),\bar{y}(\lambda ),\bar{z}(\lambda )]^\top$ are the CIE 1931 standard observer color matching functions (CMFs).

For a display with three color subpixels, there exists a unique transformation matrix $\ccmatrix \in\mathbb{R}^{3 \times 3}$ such that:
\begin{equation}
    \ccestimate(\pixscreen)=\ccmatrix\GHestimateD(\pixscreen)
    \label{eq:color_correction}.
\end{equation}

To estimate uncertainty, we need to model covariance as the color channels are no longer independent. For a three-primary (RGB) display:
\begin{equation}
    \Sigma_{\ccestimate}(\pixscreen)=\ccmatrix\Sigma_{\GHestimateD}(\pixscreen)\ccmatrix^{\top}
    \label{eq:25},
\end{equation}
where $\Sigma$ is the covariance matrix. See the supplementary material Section S2 for displays with more than three primaries.

\begin{figure*}[!t]
  \centering
      \includegraphics[width=\linewidth]{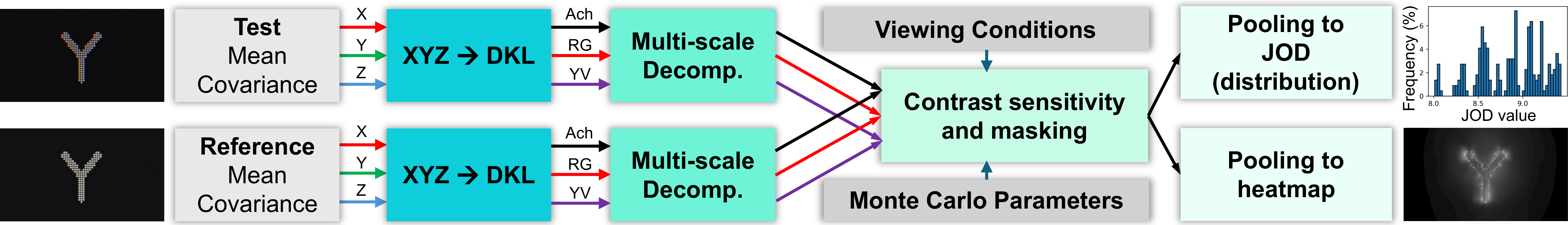}
  \caption{Visual difference predictor with uncertainty analysis. Input camera measurements (mean and covariance) are transformed to DKL space and used to generate a sample of 100 test images. The VDP pipeline evaluates the difference 21 times, each time using a different set of calibration parameters, to produce the distribution of JOD values and local differences encoded in a heatmap.} 
  \label{fig:vdp}
\end{figure*}

\subsection{Visual difference predictor}
\label{sec:visual-difference-predictor}
The visual difference predictor (VDP) is a full-reference metric operating on images represented in physical units (luminance, radiance, trichromatic values), which can estimate perceptual differences between test and reference images. We adopt the backbone architecture of ColorVideoVDP~\cite{mantiuk2024colorvideovdp} as our VDP (\figref{vdp}). The inputs are color-corrected camera measurements of the test and reference images (in XYZ space), which are first transformed into the DKL color space \edit{(color opponent space, D65 adaptation point)}. Each channel (achromatic, RG, YV) is then decomposed into spatial frequency-selective bands using a Laplacian pyramid. The contrast encoded in the bands is used to account for spatio-chromatic contrast sensitivity and contrast masking. 
\textcolor{yancheng}{Finally, the perceived contrast differences are aggregated into a JOD quality score and a corresponding heatmap, where JOD denotes Just-Objectionable-Difference. A score of 10 JOD indicates the highest quality, corresponding to identical test and reference images. The JOD scale is normalized by inter-observer variance: a decrease of 1 JOD implies that 75$\%$ of observers would detect a loss of quality in a pairwise comparison experiment.}
For further details, please refer to the original paper \cite{mantiuk2024colorvideovdp}.

The visual difference predictor involves complex computations, making analytical propagation of \edit{its aleatoric} uncertainty unfeasible. 
Therefore, we directly employ Monte Carlo sampling to estimate the uncertainties. \edit{We sample 100 test images, using the means and variances of test and reference images, which were derived from the camera measurement pipeline output ($\ccresult$, \eqref{color_correction} and \eqref{25})}. 
To estimate VDP uncertainty, we randomly partition the XR-DAVID dataset~\cite{mantiuk2024colorvideovdp} into 21 train-validation splits and train the metric for 6 epochs on each, producing 21 distinct parameter sets. Each configuration is then evaluated on 100 randomly sampled test-reference image pairs, yielding 2100 JOD estimates. The distribution of these JOD values is then treated as the outcome of uncertainty propagation through VDP.

\section{Applications and experiments}
This section first describes the implementation of camera correction, then uses Monte Carlo simulation to validate the theoretical uncertainty derivation, and finally introduces three applications: defective pixel detection, color fringing, and uniformity perception.

Measurements were conducted on an Eizo ColorEdge CS2740 display (3840$\times$2160) using a Sony $\alpha$7R III (ILCE-7RM3) camera equipped with Sony FE \textcolor{yancheng}{1.8/35mm} and Sony FE \textcolor{yancheng}{2.8/90mm} Macro G OSS lenses. To reduce Bayer demosaicing artifacts, pixel shift multi-shot mode (4-shot) was employed. All analyses were performed on RAW images (.ARQ and .ARW) processed with \textcolor{yancheng}{rawpy (v0.24.0) and libraw (v0.21.3)}. Measurements were acquired at the camera’s full resolution (7968$\times$5320). 

\begin{figure}[!t]
  \centering
      \includegraphics[width=\linewidth]{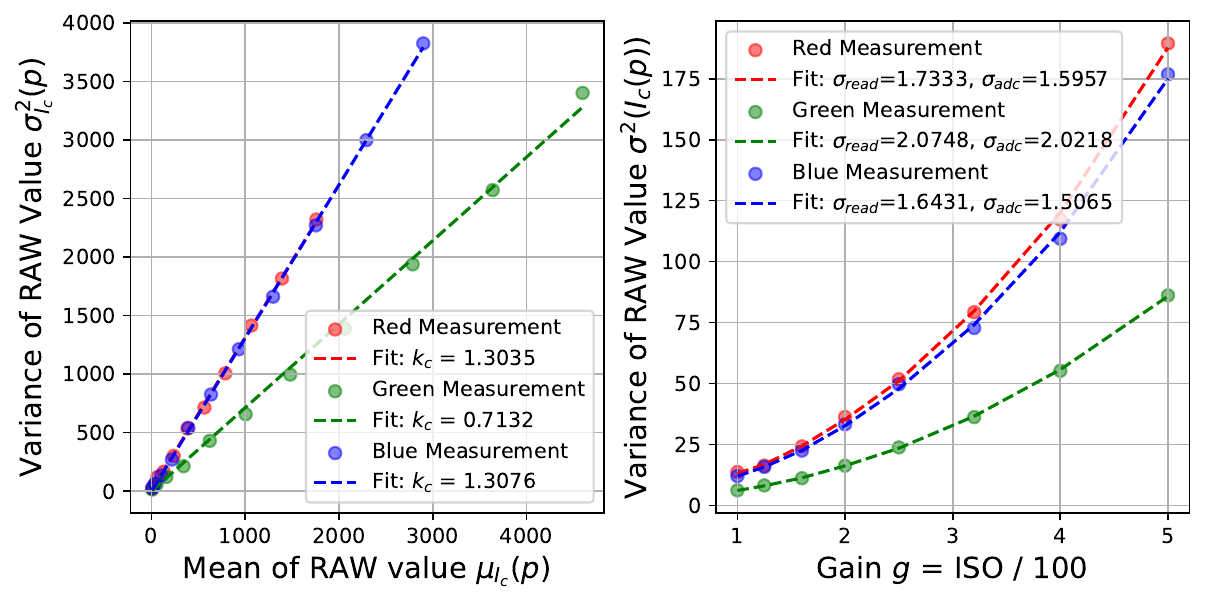}
  \caption{Noise model parameter fitting (using Sony FE \textcolor{yancheng}{2.8/90mm} lens). Left: linear fitting in bright scenes; right: quadratic fitting in dark scenes. Note that green channels are averaged across two measurements in 4-shot pixel-shift capture, so their variance is halved compared to red and blue.}
  \label{fig:noise_params}
\end{figure}

\subsection{Camera Correction Implementation}
\label{sec:camera-noise-param-est}

\paragraph{Camera noise model parameters}
The uncertainty analysis begins with~\eqref{camera-noise}, where we need to estimate $\queff_c$, $\readnoisec$, and $\adcnoisec$. From ~\eqref{camera-noise}, we have:
\begin{equation}
\sigma^2_{\rawim_c}(\pix) = \mu_{\rawim_c}(\pix) g \queff_c +\readnoisec g^{2} \queff_{c}^{2} + \adcnoisec \queff_{c}^{2}.
\label{eq:noise_paramter_estimate}
\end{equation}
Parameter estimation has two stages: (1) capture images of uniform field of varying luminance and fit the linear relationship between $\sigma^2_{\rawim_c}(\pix))$ and $\mu_{\rawim_c}(\pix)$ to estimate $\queff_c$; (2) with the camera lens covered ($\mu_{\rawim_c}(\pix)\to0$), vary $g$ (ISO/100) and fit a quadratic function between $\sigma^2_{\rawim_c}(\pix))$ and $g$ to estimate $\readnoisec$ and $\adcnoisec$. Results are in \figref{noise_params}.

\paragraph{Vignetting}
Without a uniform reference (e.g., an integrating sphere), we employed the Eizo display with uniformity correction as a flat-field source. Images were captured using a defocused lens at close range, utilizing only the display’s central region. HDR stacks were used to reduce $\epsilon_c(\pix)$, and averaging across multiple distances removed subpixel artifacts. $\hat{\vig}_c(\pix)$ are shown in~\figref{pipeline}, column 3.

\paragraph{Color Correction Matrix}
To find color correction matrix $\ccmatrix$, we measured trichromatic values of 30 colors (X-Rite 24-color ColorChecker and full/half-intensity RGB, \figref{teaser}) with JETI Specbos 1211 spectroradiometer in the center of the display. The color patches were displayed and measured sequentially. Representative color correction results (Sony FE \textcolor{yancheng}{1.8/35mm} lens + Eizo display) are shown in~\figref{pipeline}, column 6. The method achieves high accuracy, with mean color difference CIE~$\Delta E_{2000}=0.308$ between the predicted and ground-truth XYZ values.

\begin{figure}[!t]
  \centering
      \includegraphics[width=\linewidth]{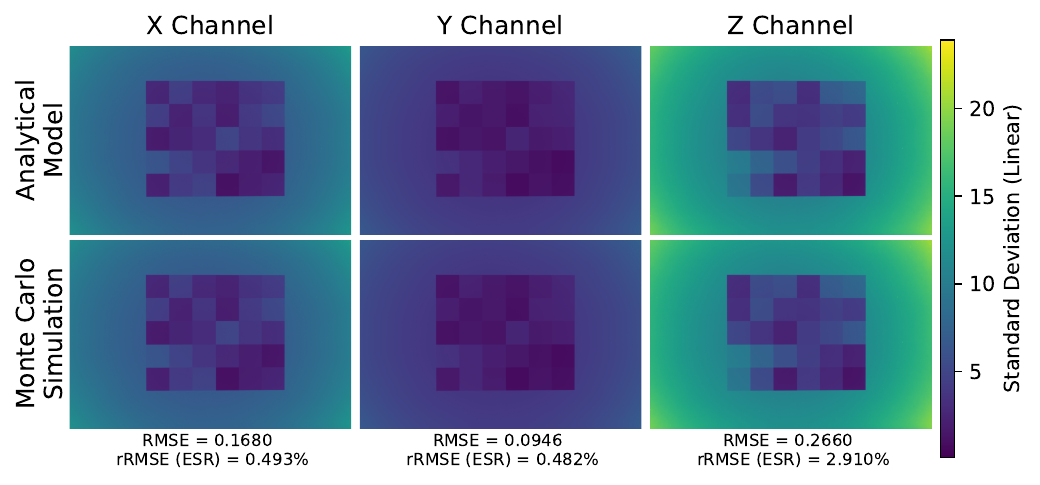}
  \caption{Monte Carlo simulations validate the analytical model of uncertainty. The top row shows results from the theoretical derivation, while the bottom row shows results of 1000 Monte Carlo samples. \textcolor{yancheng}{The RMSE quantifies the discrepancy between the analytical and Monte Carlo outputs. The rRMSE (error-to-signal ratio, ESR) is the root-mean-square of the per-pixel uncertainty error normalized by the measured value.}
}
  \label{fig:MC_full}
\end{figure}

\subsection{Monte Carlo validation of uncertainty propagation}
\label{sec:MC_Valid}
To validate our analytical uncertainty model from \secref{methodology}, we compare it with Monte Carlo simulations. Given the requirement for full-resolution data, this validation is computationally intensive. Using 1000 samples (with mean and variance of HDR-merged captures), we compare theoretical and simulated results in~\figref{MC_full}, confirming good accuracy of our uncertainty model.


\begin{table}[t]
\begin{center}
\caption{Results of display defective pixel detection and ablation study of camera measurement pipeline. The bottom row shows our complete pipeline (no color correction).  Best \textbf{PR}$_{\textcolor{yancheng}{\mathrm{AUC}}}$ under each condition is highlighted.  (HDR: HDR Image Stack Merging. MTF: MTF Inversion. VC: Vignetting Correction. GU: Geometric Undistortion. HT: Homography Transformation.)
}
\label{tab:defective_pixel}
\setlength{\tabcolsep}{1.4mm}{\scalebox{0.9}{

\begin{tabular}{c|ccc|ccc}
\toprule
Contrast   $c_{dp}$ & \multicolumn{3}{c|}{1} & \multicolumn{3}{c}{0.2} \\ \hline
Size $\dsize$ & \multicolumn{1}{c|}{4} & \multicolumn{1}{c|}{2} & 1 & \multicolumn{1}{c|}{4} & \multicolumn{1}{c|}{2} & 1 \\ \hline\hline
HDR-HT & \multicolumn{1}{c|}{0.967} & \multicolumn{1}{c|}{0.946} & 0.963 & \multicolumn{1}{c|}{0.987} & \multicolumn{1}{c|}{0.875} & 0.195 \\ \hline
HDR-MTF-HT & \multicolumn{1}{c|}{0.964} & \multicolumn{1}{c|}{0.945} & 0.978 & \multicolumn{1}{c|}{0.988} & \multicolumn{1}{c|}{0.885} & 0.252 \\ \hline
HDR-MTF-VC-HT & \multicolumn{1}{c|}{0.962} & \multicolumn{1}{c|}{0.944} & 0.976 & \multicolumn{1}{c|}{0.988} & \multicolumn{1}{c|}{0.889} & 0.213 \\ \hline
HDR-MTF-VC-GU-HT & \multicolumn{1}{c|}{\textbf{1.000}} & \multicolumn{1}{c|}{\textbf{1.000}} & \textbf{0.996} & \multicolumn{1}{c|}{\textbf{1.000}} & \multicolumn{1}{c|}{\textbf{0.908}} & \textbf{0.282} \\ \bottomrule
\end{tabular}
}}
\end{center}
\end{table}

\subsection{Detecting defective pixels on a display}
\label{sec:defective_pixel}
Prolonged use of displays may cause subpixel degradation, such as OLED burn-in. Factory defects mechanical damage can also cause defective pixels. CameraVDP precisely captures per-pixel luminance distributions and enables accurate defective pixel detection.

\paragraph{Test patterns}
Test patterns simulate a display with defective pixels. Each pattern consists of a uniform white image (pixel value [255, 255, 255]) displayed on a $3840\times2160$ screen, containing 100 darker square patches simulating defective pixels. All patches within a stimulus share the same edge length $\dsize\in\{1,2,4\}$ (display pixels) and Weber contrast $\dcontrast\in\{0.2,1\}$ (corresponding to sRGB-encoded pixel values of 231 and 0) relative to the background. Varying $\dsize,\dcontrast$ yields 6 unique test patterns. Experiments were conducted using the Eizo display, the Sony $\alpha$7R III camera with the FE \textcolor{yancheng}{1.8/35mm} lens.
\paragraph{Detection method}
Given parameters $\dsize,\dcontrast$, display size $\dissizew \times \dissizeh$ (display pixels), and the oversampling factor $o$ used in homography transformation, the measurement resolution is 
\textcolor{yancheng}{$(o\dissizew +1 )\times (o\dissizeh+1)$}
(image pixels). The mean map $\meanmap$ is computed by averaging the RGB channels of the output from~\eqref{remapping} (no color correction), then convolving with a box filter of size \textcolor{yancheng}{$(o\,\dsize-1)$}. Similarly, a background mean map $\backmap$ is obtained using a box filter size of \textcolor{yancheng}{$(10o+1)$}. A region is labeled defective if $\meanmap < \backmap \dthreshold$, where detection threshold $\dthreshold \in [0,1]$. Redundant detections are removed using non-maximum suppression (NMS)~\cite{canny1986computational}.

\paragraph{Evaluation metrics}
The objective is to identify defective pixels with minimal number of false positives and negatives. The evaluation metric is the area under the precision-recall curve (\textbf{PR}$_{\textcolor{yancheng}{\mathrm{AUC}}}$), computed across varying detection thresholds $\dthreshold$. A detection is considered correct if the predicted center lies within a one-display-pixel radius of the ground truth center.

\paragraph{Results}
Experimental results in the bottom row of \tableref{defective_pixel} show that for $\dsize \geq 2$, our method achieves near-perfect $\mathbf{PR}_{\textcolor{yancheng}{\mathrm{AUC}}}$ ($\approx$1). Under extreme conditions ($\dsize = 1,\dcontrast = 0.2$), $\mathbf{PR}_{\textcolor{yancheng}{\mathrm{AUC}}}$ decreases due to noise (more false positives) as discussed below. Thus, we recommend placing the camera closer to the display to enlarge low-contrast defective pixels in practical use.
\paragraph{Ablation studies}
The contribution of each component of our camera correction pipeline to the accuracy of detection is reported in~\tableref{defective_pixel}. The baseline HDR-HT method showed the poorest performance. Incorporating the MTF inversion module (Wiener deconvolution) in HDR-MTF-HT significantly improved results for small defect sizes, confirming the MTF module’s effectiveness and highlighting glare and blur as key challenges. Vignetting correction (VC) had negligible impact since detection uses relative local background to compensate for vignetting. Geometric undistortion (GU) enabled the full pipeline to achieve near-perfect performance on larger defect sizes.

\begin{figure}[!t]
  \centering
      \includegraphics[width=\linewidth]{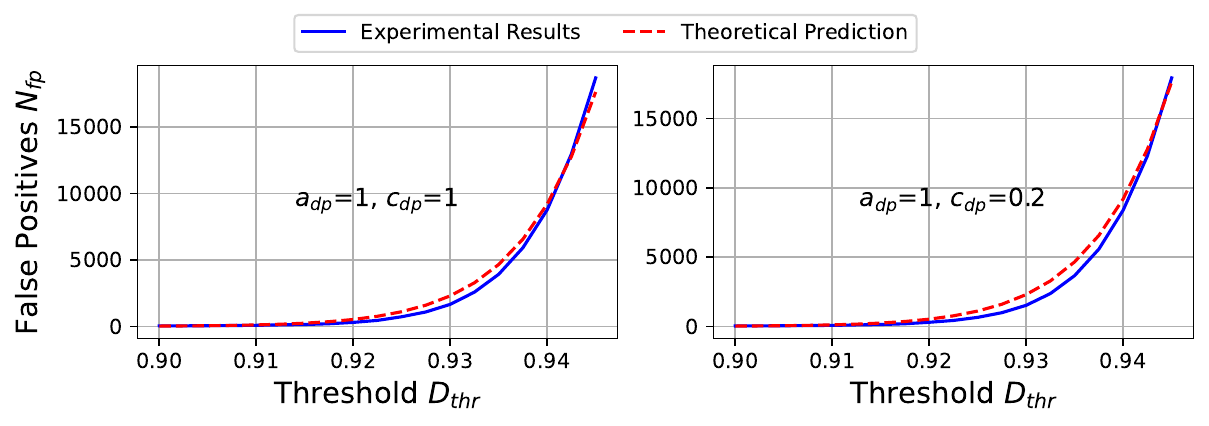}
  \caption{Estimating false positive number via uncertainty map for trustworthy detection without any prior knowledge of defective pixel conditions.; blue: empirical, red dashed: theoretical from uncertainty; left/right: different contrast corruptions (contrast does not affect false positive number). Only $D_{\text{thr}}\in[0.900,0.945]$ is shown—smaller values yield near-zero false positives, while larger values exceed the computational limits of NMS.
}
  \label{fig:predict_dp}
\end{figure}
\paragraph{Uncertainty (Noise) Model} 
In practice, ground truth is unavailable—defect pixel number, location, and contrast are unknown. A high detection threshold $D_{\text{thr}}$ (near 1) risks excessive false positives, while a low threshold may miss low-contrast defects. Our uncertainty map can be used to estimate a false positive count $\nfp$, offering a reliability metric for detection outcomes.

Since $\meanmap$ already contains noise, we use $\backmap$ (computed with a larger box filter and thus assumed to be less affected by noise) for false positive estimation. We consider only pixels with $\dsize = 1$ and estimate the probability that each \textbf{display} pixel $\pixscreen$ with $\mu(\pixscreen) = \backmap(\pixscreen), \sigma^2(\pixscreen) = \frac{\sum_{c} \sigma_c^2(\GHestimate(\pixscreen))}{9}$ satisfies condition $\meanmap'(\pixscreen) < \backmap(\pixscreen) \dthreshold$ (false positive). Note that $\meanmap'(\pixscreen)$ is a sample drawn from normal distribution with $\mu(\pixscreen)$ and $\sigma^2(\pixscreen)$, and is independent of $\meanmap(\pixscreen)$. Then, for pixel $\pixscreen$, the z-score is given by:
\begin{equation}
    z(\pixscreen)=\frac{\left(\dthreshold-1\right) \backmap(\pixscreen)}{\sigma(\pixscreen)}
    \label{eq:z-score}
\end{equation}
We can then compute the cumulative distribution function (CDF), representing the probability that the pixel becomes a false positive:
\begin{equation}
    \Phi(z(\pixscreen))=\frac{1}{2}\left(1+\operatorname{erf}\left(\frac{z(\pixscreen)}{\sqrt{2}}\right)\right)
    \label{eq:CDF},
\end{equation}
where $\operatorname{erf}$ is the Gauss error function. The final estimate of the number of false positive detections is:
\begin{equation}
    \nfp = \sum_{\pixscreen} \Phi(z(\pixscreen))
    \label{eq:NFP},
\end{equation}
Experimental results are shown in~\figref{predict_dp}. Without any ground truth, our method accurately predicts the number of false positives and, therefore, can be used to select the desired detection threshold.

\begin{figure}[!t]
  \centering
      \includegraphics[width=\linewidth]{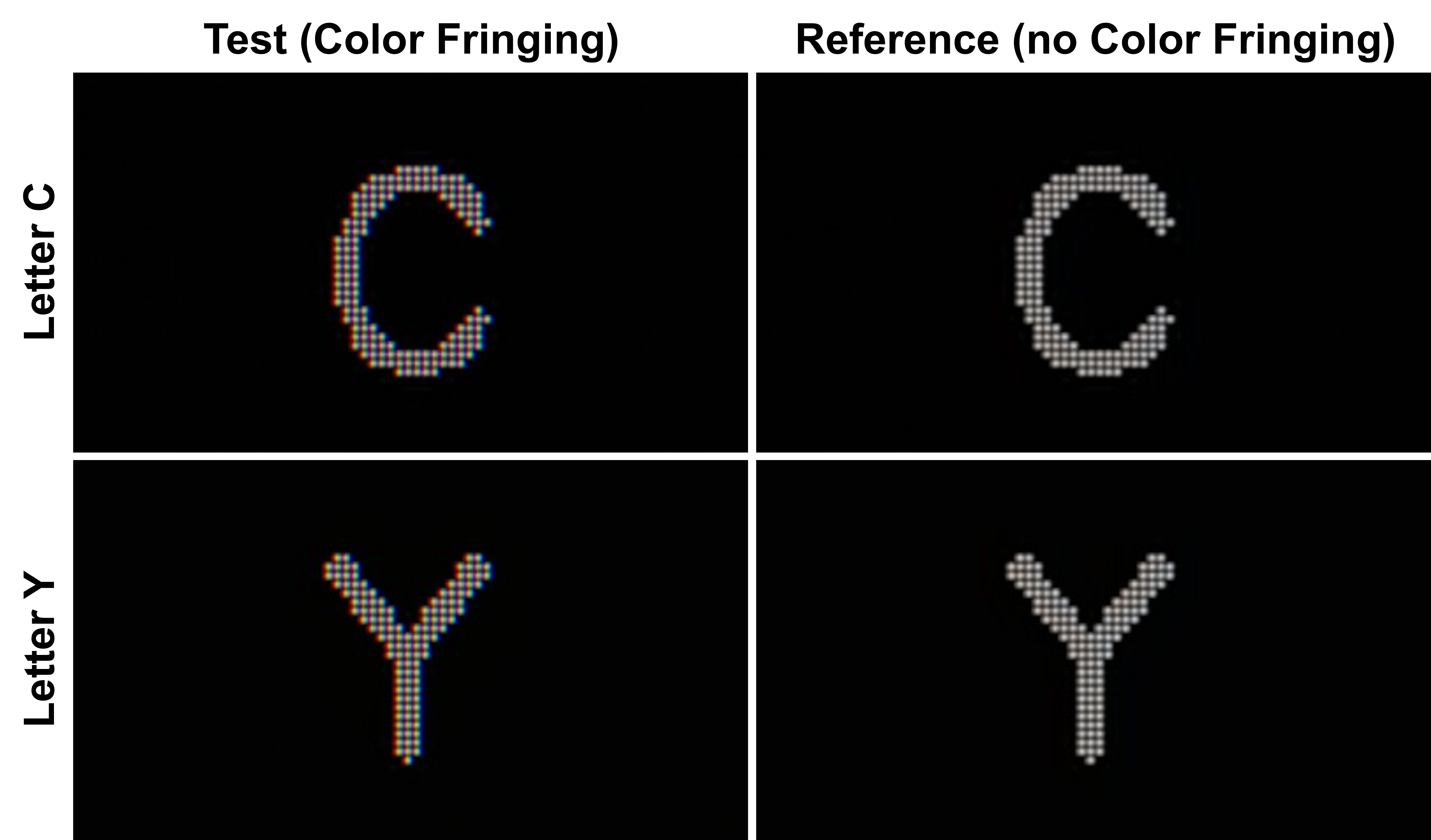}
  \caption{Test and reference images for color fringing psychophysical experiments. The images have been center-cropped at 10× magnification to facilitate comparison; the actual stimuli used in the experiments were significantly smaller. Note that due to variations in display primaries, the test (left) and reference (right) images may appear chromatically different on some displays, although they are perceptually matched on the Eizo display.}
  \label{fig:color_fringing_example}
\end{figure}
\begin{figure}[!t]
  \centering
      \includegraphics[width=\linewidth]{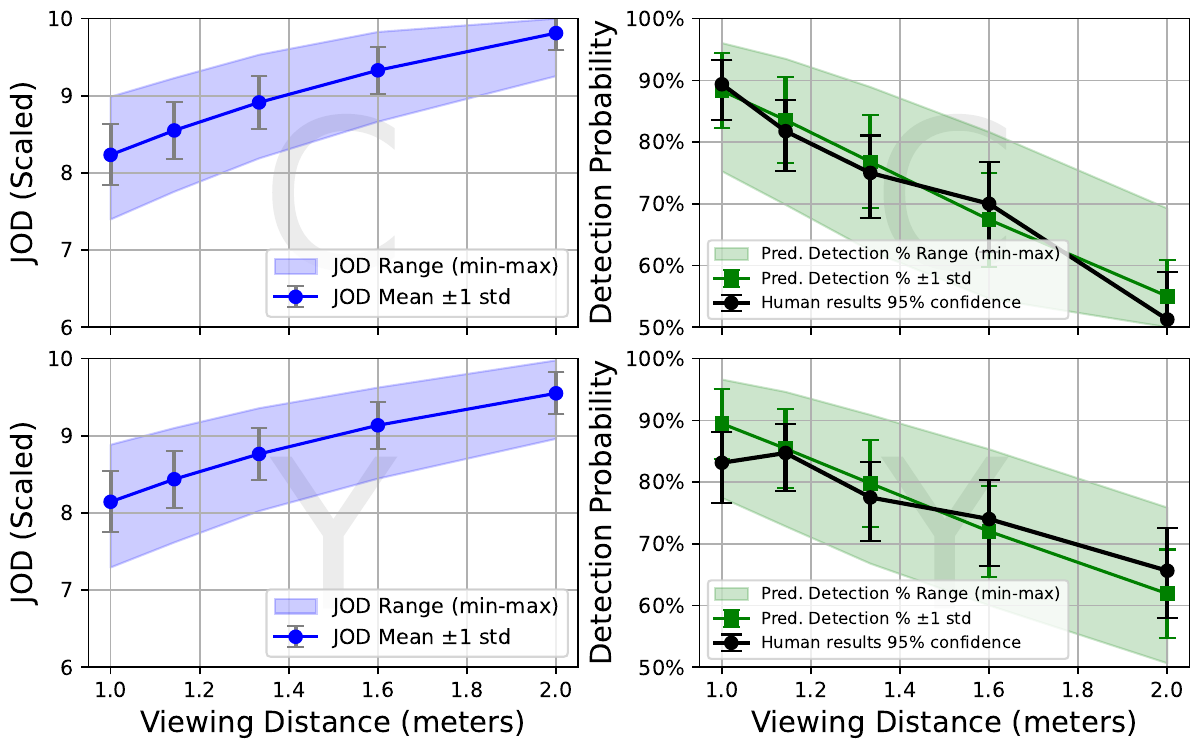}
  \caption{Color fringing experiment results. Top: letter C; bottom: letter Y. Left: the scaled JOD values predicted by the visual difference predictor, along with standard deviation (error bars) and extrema (shader region) as measures of uncertainty. Right: the human experimental results (mean and 95\% confidence intervals) in black, and the predicted detection probability (mean, standard deviation, and extrema converted from JOD). The predictions closely match human performance.}
  \label{fig:color_fringing_result}
\end{figure}
\subsection{Display color fringing}
\label{sec:color_fringing}
RGB subpixel layouts often introduce color fringing artifacts like red-blue edges around white text. This section evaluates whether CameraVDP can accurately predict the visibility of such artifacts.
\paragraph{Stimulus}
To accurately reproduce color fringing, we displayed the letters``C'' and ``Y'' on the Eizo display and captured them with the Sony $\alpha$7R III and FE \textcolor{yancheng}{2.8/90mm} lens. The experiment included a test image (with color fringing) and a reference (without fringing). Because the pixel density of our display was too high to see color fringing, we simulated a lower resolution display (e.g., VR headset) by enlarging the captured image 4× in both dimensions. The reference image was generated by converting the test image to grayscale. Test and reference images are shown in~\figref{color_fringing_example}.
\paragraph{Experimental procedure}
We investigated the effect of viewing distance on color fringing visibility using a two-alternative forced choice (2AFC) protocol. Stimuli were displayed on the Eizo monitor mounted on a motorized movable rail~\cite{ashraf2025}. In each trial, the display was moved to one of five viewing distances [1.00, 1.14, 1.33, 1.60, 2.00] meters, and the corresponding pixel-per-degree resolutions (with the simulated 4×4 pixels) are [28.9, 32.8, 38.0, 45.4, 56.5]. Each condition was presented 20 times per observer in random order (of distances and ``C''/``Y'' letters). Each trial consisted of two intervals, one with color fringing (test) and one without (reference), each lasting 2\,s, separated by 0.5\,s of random noise. Observers were asked to select the interval containing color fringing. 
\paragraph{Participants}
We recruited 8 participants (5 male, 3 female) aged 20–35, all of whom passed a color vision and acuity screening. The experiment was approved by the departmental ethics committee.
\paragraph{Results and discussion}
The results are shown in~\figref{color_fringing_result}, including both human data and model predictions. To account for display characteristics, the test and reference patterns were re-captured using the Sony $\alpha$7R III camera with the FE \textcolor{yancheng}{1.8/35mm} lens before being processed by the visual difference predictor to obtain JOD values, which were then converted to detection probabilities~\footnote{Since the VDP was trained on full-field images, but color fringing stimuli are very small, JOD values were rescaled as $\text{JOD}_{\text{scale}} = \min(11.35 - 20 \times (10-\text{JOD}), 10)$.}. These constitute the model predictions, shown as purple and green lines (mean) and shaded areas (uncertainty) in~\figref{color_fringing_result}. Human observer data are black lines in the right panel. The results demonstrate that our pipeline accurately predicts human detection probabilities of color fringing artifacts, with model uncertainty encompassing the 95\% confidence interval of human results.

\begin{figure}[!t]
  \centering
      \includegraphics[width=\linewidth]{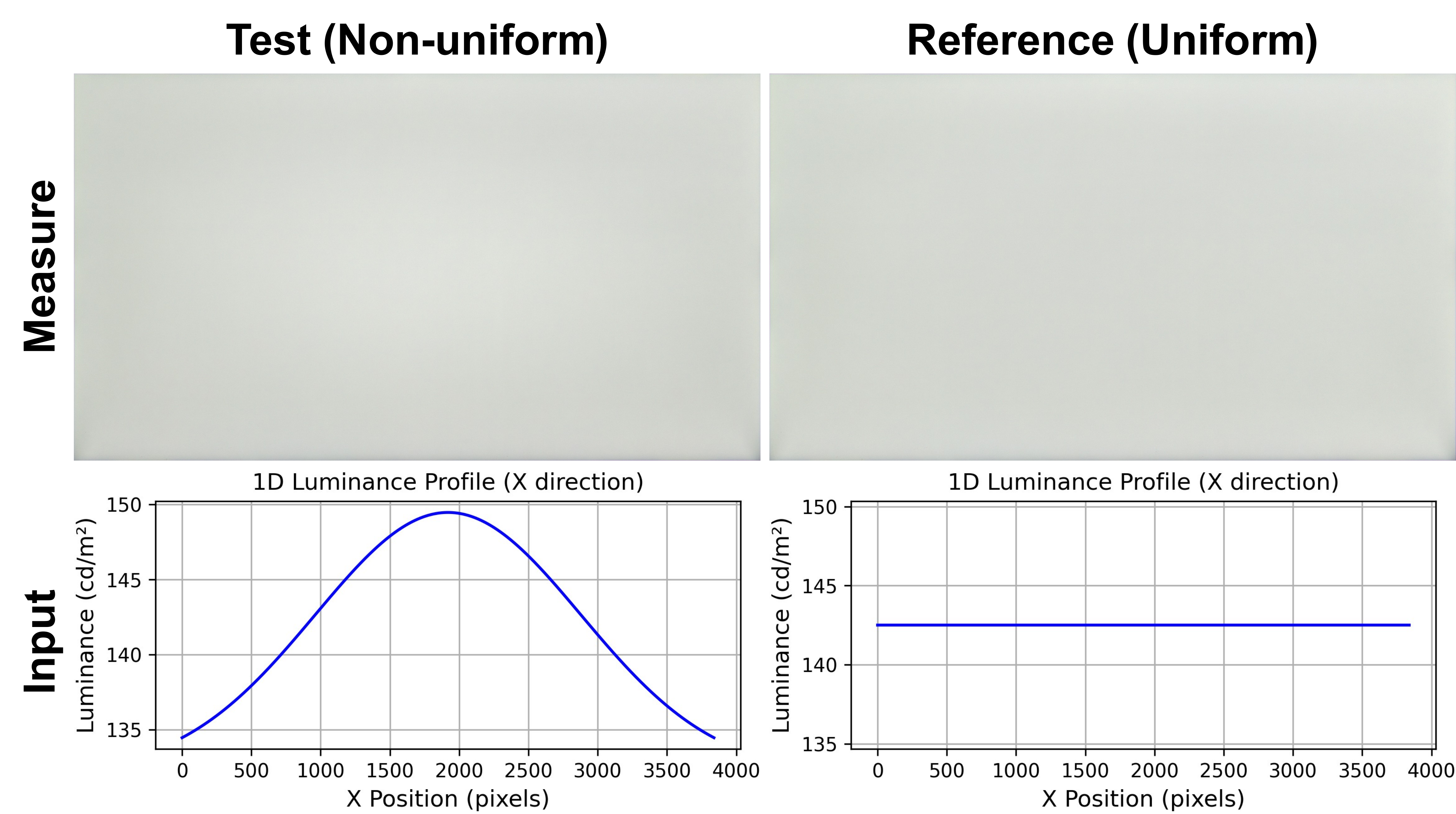}
  \caption{Test and reference images for display uniformity assessment psychophysical experiments. The first row shows the display output measured by Sony FE \textcolor{yancheng}{1.8/35mm} (converted to sRGB), and the second row shows the 1D luminance profile of the input signal along the X-axis. Reducing the page scale will make the central bright spot in the test stimulus more prominent.
}
  \label{fig:uniformity}
\end{figure}
\begin{figure}[!t]
  \centering
      \includegraphics[width=\linewidth]{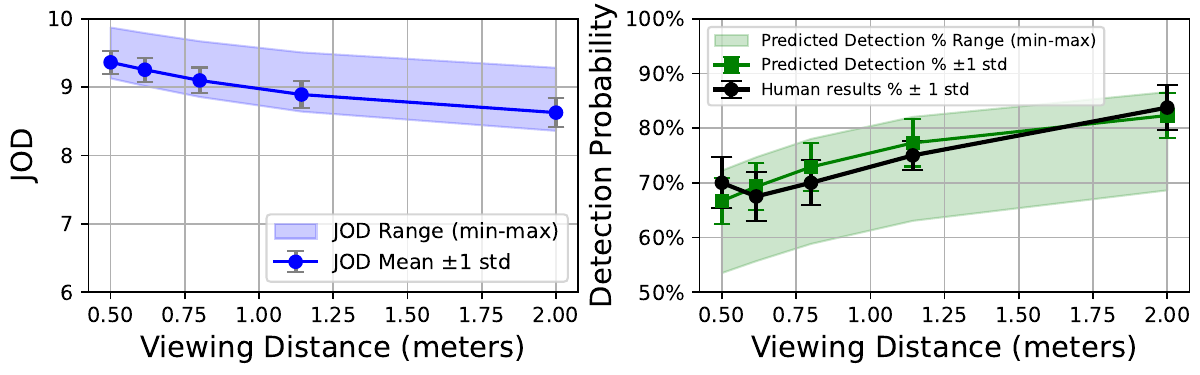}
  \caption{Display non-uniformity assessment experiment results. The notation is the same as in \figref{color_fringing_result}.}
  \label{fig:uniformity_results}
\end{figure}
\subsection{Display non-uniformity assessment}
\secref{defective_pixel} examined high-frequency non-uniformities (defective pixels), whereas low-frequency artifacts (e.g., center-bright vignetting from backlight unevenness) are also common. This section evaluates whether CameraVDP can accurately predict their visibility as a function of viewing distance.
\paragraph{Stimulus}
We pre-measured the Eizo display’s non-uniformity $\frac{L_{\max}-L_{\min}}{L_{\max}+L_{\min}}=1.94\%$, which was sufficient for the experiment.
The test stimulus was a 2D elliptical Gaussian with a contrast of 0.1 and a mean of 142.5\csdm, with horizontal and vertical standard deviations set to half of the width and height, as shown in~\figref{uniformity}. The reference was a uniform field matched in mean luminance. The display was driven by a 12-bit per color channel signal (10-bit native + 2 bits simulated with spatio-temporal dithering) to avoid banding.
\paragraph{Experimental procedure and participants}
The experimental procedure and participants were nearly identical to those in~\secref{color_fringing}, with two exceptions: (1) viewing distances of [0.50, 0.62, 0.80, 1.14, 2.00] meters were used; (2) to prevent detection based on luminance differences rather than non-uniformity, the reference luminance was randomly scaled by a factor in the range from 0.8 to 1.2.
\paragraph{Results and discussion}
The results are presented in~\figref{uniformity_results}. Overall, the findings indicate that display non-uniformities are generally perceived more strongly at greater viewing distances, as the increase in viewing distance corresponds to higher spatial frequencies, to which the human visual system is more sensitive (see~\figref{uniformity}; this can also be observed by zooming the page in and out). CameraVDP accurately predicted this effect.

\section{Conclusions}
A consumer-grade camera can become an accurate measurement instrument when properly calibrated. Here, we calibrate the camera specifically for the task of capturing a display, so that we can take advantage of the pixel grid to refine geometric calibration and ensure accurate color measurement. Because not every display artifact is going to be visible, we integrate our camera correction pipeline with a visual difference predictor. Finally, our measurements come with an estimate of uncertainty, which is introduced by sensor noise and VDPs prediction error. Our measurement technique can be used across a range of (computational) display applications, from detecting defects to evaluating the visibility of display artifacts. 
\subsubsection*{Limitations}
CameraVDP has \textcolor{yancheng}{four} main limitations: (1) it is applicable only to cameras that are good enough to be calibratable; (2) it requires cameras with pixel-shift capability, which does not rely on demosaicing for color reconstruction. Demosaicing may introduce inter-pixel dependencies that violate the assumption of statistical independence; (3) the increase of uncertainty due to the MTF inversion is approximated assuming white noise, which may not hold for scenes with large variation of intensity; \textcolor{yancheng}{(4) our experimental validation was conducted on a single display.}


\begin{acks}
We would like to thank Dongyeon Kim for the helpful discussion and advice. We are grateful to anonymous reviewers for their feedback. 
\end{acks}

\bibliographystyle{ACM-Reference-Format}
\bibliography{Camera_Display,refs}
\appendix
\clearpage

\end{document}


\newcommand{\figref}[1]{Figure~\ref{fig:#1}}
\newcommand{\secref}[1]{Section~\ref{sec:#1}}
\newcommand{\algoref}[1]{Algorithm~\ref{algo:#1}}
\newcommand{\chapterref}[1]{Chapter~\ref{chapter:#1}}
\newcommand{\appref}[1]{Appendix~\ref{app:#1}}
\newcommand{\tableref}[1]{Table~\ref{tab:#1}}
\newcommand{\etal}{et al.\xspace}
\newcommand{\degpers}{\,\nicefrac{deg}{s}\xspace}
\newcommand{\degperssq}{\,\nicefrac{deg}{s\textsuperscript{2}}\xspace}
\newcommand{\degree}{$^{\circ}$\xspace}
\newcommand{\Hz}{\,Hz\xspace}
\newcommand{\csdm}{\,cd/m$^2$}
\newcommand{\fourier}{\mathfrak{F}}
\newcommand{\infourier}{^{\mathfrak{F}}}

\LetLtxMacro{\originaleqref}{\eqref}
\renewcommand{\eqref}[1]{Eq.~\originaleqref{eq:#1}}

\newcommand{\todo}[1]{\textcolor{red}{\textbf{todo: #1}}}
\newcommand{\RM}[1]{\textcolor{brown}{\textnormal{(Rafal) #1}}}
\newcommand{\AB}[1]{\textcolor{purple}{\textnormal{(Ali) #1}}}
\newcommand{\MA}[1]{\textcolor{blue}{\textnormal{(Maliha) #1}}}
\newcommand{\MAtext}[1]{\textcolor{blue}{\textnormal{#1}}}

\newcommand{\edit}[1]{\textcolor{black}{#1}}

\definecolor{yancheng}{RGB}{0,0,0}

\newcommand{\ourmethod}{elaTCSF}

\newcommand{\code}[1]{\texttt{#1}}

\newcommand{\ind}[1]{\text{#1}}


\title{CameraVDP: Perceptual Display Assessment with Uncertainty Estimation via Camera and Visual Difference Prediction - Supplementary Materials}

\author[]{Yancheng Cai}
\email{yc613@cam.ac.uk}
\affiliation{
  \institution{University of Cambridge}
  \streetaddress{William Gates Building, 15 JJ Thomson Avenue}
  \city{Cambridge}
  \postcode{CB3 0FD}
  \country{United Kingdom}
}

\author[]{Robert Wanat}
\email{robwanat@gmail.com}
\affiliation{
  \institution{LG Electronics North America}
  \city{Santa Clara}
  \country{United States of America}
}

\author[]{Rafał K. Mantiuk}
\email{rafal.mantiuk@cl.cam.ac.uk}
\affiliation{
  \institution{University of Cambridge}
  \streetaddress{William Gates Building, 15 JJ Thomson Avenue}
  \city{Cambridge}
  \postcode{CB3 0FD}
  \country{United Kingdom}
}


\begin{abstract}
This supplementary material includes: (1) a table of all variables and their definitions, (2) a complete derivation of uncertainty propagation in MTF inversion, and (3) the computation of mean and uncertainty in color correction for multi-primary displays (e.g., WRGB OLEDs).
\end{abstract}

\maketitle

\section*{Content}

\begin{itemize}
    \item Section \ref{sec:uncertainty_MTF} --- Uncertainty propagation in MTF inversion
    \item Section \ref{sec:uncertainty_color_correction} --- Color Correction for 4+ primaries
    \item Table \ref{tab:symbols} --- List of symbols
    \item Table \ref{tab:parameters} --- Parameter fitting results
\end{itemize}

\begin{table*}[tbp]
\footnotesize
\centering
\caption{Table of key symbol definitions used in the main text.}
\label{tab:symbols}
\renewcommand{\arraystretch}{1.3}
\begin{tabularx}{\textwidth}{lX}
\textbf{Variable Symbols} & \textbf{Definition} \\
\hline
$\dsize$ & Side length of the simulated defective pixel in the defective pixel detection experiment, expressed in units of display pixel size. \\
$c$ & Color channels in the camera RAW image, $c \in \{\mathsf{r},\mathsf{g},\mathsf{b}\}$. \\
$\dcontrast$ & Weber contrast of the simulated defective pixel. \\
$C_c$ & The spectral sensitivity of camera channel $c \in \{\mathsf{r},\mathsf{g},\mathsf{b}\}$. \\
$\dark$ & Camera dark noise. \\
$\dthreshold$ & Defective pixel detection threshold, $\dthreshold \in [0,1]$. \\
$E_{\dissubp}$ & Spectral power distribution (SPD) of display subpixels, $\mathsf{k} \in \{\mathsf{R}, \mathsf{G}, \mathsf{B}\}$. \\
$\mathcal{F}$ & Fourier transform. \\
$g$ & Camera gain. \\
$\pixundistort$ & Camera pixel coordinates in undistorted images, with the image center as the origin; x-axis pointing right, y-axis pointing upward. \\
$G$ & Wiener filter. \\
$h$ & Coordinate mapping of the homography transformation, $h: \pixscreen \to \pixundistort$. \\
$\homo$ & Homography matrix. \\
$i$ & Each RAW exposure in HDR image stack. \\
$\mathbf{I}$ & RAW (digital) sensor values captured by the camera, represented as a matrix of linear values. \\
$J$ & Jacobian matrix used in the uncertainty propgation for display with more than three subpixels. \\
$\dissubp$ & Display subpixels, $\mathsf{k} \in \{\mathsf{R}, \mathsf{G}, \mathsf{B}\}$. \\
$\queff_c$ & Camera quantum efficiency of camera color channels. \\
$\noMTFrad$ & Clean scene radiance (result of MTF inversion). \\
$\VCradc$ & Result of vignetting correction. \\
$\GHD$ & Result of geometric transformation. \\
$m$ & Global coordinate mapping of the geometric transformation, $m: \pixscreen \to \pix$. \\
$M(\freq)$ & Modulation transfer function (MTF). \\
$\backmap$ & Background mean map in the defective pixel detection experiment. \\
$\meanmap$ & Mean (pixel surrounding) map in the defective pixel detection experiment. \\
$\ccmatrix$ & Color correction matrix for display with three subpixels. \\
$\ccmatrixroot$ & Color correction matrix for display with more than three subpixels. \\
$N$ & Total number of RAW images in the HDR image stack. \\
$N_{(.)}(\freq)$ & The power spectral densities (PSD) of the noise. \\
$\mathbb{N}_{(.)}(\pix)$ & Noise component of the variable (zero mean). \\
$o$ & Display pixel oversampling factor. \\
$\pix$ & Camera pixel coordinates in RAW images, with the image center as the origin; x-axis pointing right, y-axis pointing upward. \\
$P_{\dissubp}$ & Linearized display subpixel value, $\mathsf{k} \in \{\mathsf{R}, \mathsf{G}, \mathsf{B}\}$. \\
$\psf$ & Camera point spread function (PSF). \\
$\pixscreen$ & Display pixel coordinates in Display panel, with the display center as the origin; x-axis pointing right, y-axis pointing upward. \\
$S_{(.)}(\freq)$ & The power spectral densities (PSD) of the signal. \\
$\mathbb{S}_{(.)}(\pix)$ & Signal component of the variable (zero variance). \\
$\bar{\mathbf{S}}$ & The CIE 1931 standard observer color matching functions (CMFs). \\
$t$ & Camera exposure time, in seconds (s). \\
$u$ & Coordinate mapping of the geometric undistortion, $u: \pixundistort \to \pix$. \\
$\vig$ & Vignetting function. \\
$x$ & A single measurement of $\rerad$. \\
$\rerad$ & Relative radiance, an intermediate variable for computing $\scrad$. \\
$\ccresult$ & Result of color correction, in XYZ color space. \\
$\epsilon$ & Noise in Vignetting correction. \\
$\eta$ & Noise in MTF inversion. \\
$\lambda$ & The wavelength. \\
$\mu_{(.)}$ & Mean of the normal distribution. \\
$\sigma^2_{(.)}$ & Variance of the normal distribution. \\
$\Sigma_{(.)}$ & Covariance matrix of the multivariate normal distribution. \\
$\adcnoise$ & Camera analog-to-digital conversion noise (ADC noise). \\
$\readnoise$ & Camera read noise. \\
$\scrad$ & Scene radiance, the reconstruction target in the HDR image stack merging section. \\
$\freq$ & Spatial frequency in cycles-per-pixel. \\
$\mathrm{Pois}$ & Poisson distribution. \\
$\mathcal{N}$ & Normal distribution. \\
\end{tabularx}
\end{table*}

\begin{table}[t]
\begin{center}
\caption{Parameter fitting results.}
\label{tab:parameters}
\setlength{\tabcolsep}{4mm}
\renewcommand{\arraystretch}{1.3} 
\scalebox{1}{
\begin{tabular}{c|c}
\toprule
Part & Parameters \\ \hline\hline
Noise Model & \begin{tabular}[c]{@{}c@{}} 
$\queff_c$ = {[}1.303514, 0.713188, 1.307612{]},\\
$\readnoise$ = {[}1.733335, 2.074783, 1.643126{]},\\
$\adcnoise$ = {[}1.595734, 2.021769, 1.506513{]}
\end{tabular} \\ \hline
MTF & \begin{tabular}[c]{@{}c@{}} 
$a_1$ = 0.00174, $b_1$ = 0.67193, $c_1$ = 0.12362,\\
$a_2$ = 1.30353, $b_2$ = -0.11405, $c_2$ = 0.22962
\end{tabular} \\ 
\bottomrule
\end{tabular}
}
\end{center}
\end{table}

\section{Uncertainty propagation in MTF inversion}
\label{sec:uncertainty_MTF}
In the main text, we provide only a brief derivation and result of the MTF inversion. Here, we present the complete derivation and proof.

The scene radiance $\scrad_c(\pix)$ affected by blur and glare can be modeled as the convolution of the clean scene radiance $\noMTFradc(\pix)$ and the camera point spread function (PSF) $\psf(\pix)$, plus noise $\eta(\pix)$:
\begin{equation}
    \scrad_c (\pix) = (\noMTFradc * \psf)(\pix) + \eta(\pix)
    \label{eq:MTF_1},
\end{equation}
where $*$ denotes convolution. The MTF $M(\freq)$ is the modulus of the Fourier transform of the PSF $\psf\pix)$~\cite{burns2022updated}, $\freq$ is spatial frequency in cycles-per-pixel. Taking the Fourier transform $\mathcal{F}$ on both sides of~\eqref{MTF_1} yields:
\begin{equation}
    \mathcal{F} (\scrad_c) = \mathcal{F} (\noMTFradc) M(\freq) + \eta'(\freq)
    \label{eq:MTF_2},
\end{equation}
Considering the noise, we use Wiener deconvolution to obtain the deglared and deblurred estimate of $\noMTFradc$:
\begin{equation}
\noMTFradcestimate(\pix) = \mathcal{F}^{-1}\left(\mathcal{F}\left(\hat{\scrad}_c\right) G_c(\freq)\right)(\pix)
    \label{eq:MTF_3}\,,
\end{equation}
where the Wiener filter is ($^{*}$ is the conjugate operator):
\begin{equation}
G_c(\freq)=\frac{M^{*}(\freq) S_{\hat{\scrad}_c}(\freq)}{|M(\freq)|^{2} S_{\hat{\scrad}_c}(\freq)+N_{\hat{\scrad}_c}(\freq)}
    \label{eq:MTF_4}\,.
\end{equation}
$S_{\hat{\scrad}_c}(\freq)$ and $N_{\hat{\scrad}_c}(\freq)$ represent the power spectral densities (PSD) of the signal $\mathbb{S}_{\hat{\scrad}_c}(\pix)$ and noise $\mathbb{N}_{\hat{\scrad}_c}(\pix)$:
\begin{equation}
    \hat{\scrad}_c(\pix) = \mathbb{S}_{\hat{\scrad}_c}(\pix) + \mathbb{N}_{\hat{\scrad}_c}(\pix)
    \label{eq:MTF_5}\,,
\end{equation}
\begin{equation}
    \mathbb{S}_{\hat{\scrad}_c}(\pix) \sim N\left(\mu_{\hat{\scrad}_c}(\pix),0\right)
    \label{eq:MTF_6}\,,
\end{equation}
\begin{equation}
    \mathbb{N}_{\hat{\scrad}_c}(\pix) \sim  N\left(0,\sigma^2_{\hat{\scrad}_c}(\pix)\right)
    \label{eq:MTF_7}\,.
\end{equation}
Thus $S_{\hat{\scrad}_c}(\freq)$ can be directly computed:
\begin{equation}
    S_{\hat{\scrad}_c}(\freq) = \left|\mathcal{F}\left(\mathbb{S}_{\hat{\scrad}_c}\right)\right|^2 = \left|\mathcal{F}\left(\mu_{\scrad_c}\right)\right|^2
    \label{eq:MTF_8}\,.
\end{equation}
However, since $\mathbb{N}_{\hat{\scrad}_c}(\pix)$ is a random variable and not directly accessible, the computation of $N_{\hat{\scrad}_c}(\freq)$ differs accordingly. The noise $\mathbb{N}_{\hat{\scrad}_c}(\pix)$ is assumed to be white in the spatial frequency domain (i.e., equal magnitude in all frequencies). According to Parseval's theorem, the energy of the noise signal in the spatial domain equals that in the frequency domain, so:
\begin{equation}
    N_{\hat{\scrad}_c}(\freq) \approx  \overline{ \sigma^2_{\hat{\scrad}_c}(\pix)} 
    \label{eq:MTF_9}\,.
\end{equation}
For simplicity, the MTF was assumed to be isotropic. We measured \(M(\freq)\) using the slanted-edge method, as shown in the second column of Figure~M\ref{main-fig:pipeline} (main text). Specifically, images were captured of the diagonal black-white edge at the center of a Siemens star rotated by \(45^\circ\). The MTF was fitted using: 
\begin{equation}
M'(\freq)=a_{1} \exp \left(-\left(\frac{\freq-b_{1}}{c_{1}}\right)^{2}\right)+a_{2} \exp \left(-\left(\frac{\freq-b_{2})}{c_{2}}\right)^{2}\right)\,,   
\label{eq:MTF_10}
\end{equation}
where \(a_1, a_2, b_1, b_2, c_1, c_2\) are fitting parameters. To suppress noise amplification, $M(\freq) = \max(M'(\freq), 0.5)$, as shown in Figure~M\ref{main-fig:ESF_MTF_results}.

Since the Fourier transform is a linear operation, $\hat{L}_{c}(\pix)$ follows a normal distribution. The mean of $\noMTFradcestimate(\pix)$ is given by~\eqref{MTF_3}:
\begin{equation}
    \mu_{\noMTFradcestimate}(\pix) = \mathcal{F}^{-1}\left(\mathcal{F}\left(\mu_{\scrad_c}\right) G(\freq)\right)(\pix)
    \label{eq:MTF_11}\,,
\end{equation}

Now we aim to compute the variance $\sigma^2(\noMTFradcestimate(\pix))$. Our analysis relies on the following assumptions: (1) the camera PSF $\psf$ decays rapidly in spatial domain; (2) the noise follows a white noise distribution; and (3) local variance is approximately uniform within small pixel neighborhoods. These assumptions hold for most well-focused cameras and typical scenes.

Revisiting~\eqref{MTF_1}, since $\psf$ decays rapidly in spatial domain, the following expression still holds within the neighborhood of pixel $\pix$:
\begin{equation}
    \scrad_c’(\pix) = (\noMTFradcp * \psf)(\pix) + \eta(\pix)
    \label{eq:MTF_12},
\end{equation}
where $'$ indicates values within the neighborhood of pixel $\pix$. 

From~\eqref{MTF_3}, we have:
\begin{equation}
    \mathcal{F}\left(\noMTFradcpestimate\right)(\freq)=\left(\mathcal{F}\left(\hat{\scrad}'_{c}\right)G\right)(\freq)
    \label{eq:MTF_13}\,.
\end{equation}
Decompose each component into the sum of signal and noise:
\begin{equation}
    \mathcal{F}\left(\mathbb{S}_{\noMTFradcpestimate} + \mathbb{N}_{\noMTFradcpestimate}\right)(\freq)=\left(\mathcal{F}\left(\mathbb{S}_{\hat{\scrad}'_c}+\mathbb{N}_{\hat{\scrad}'_c}\right)G\right)(\freq)
    \label{eq:MTF_14}\,.
\end{equation}
Due to the linearity of the Fourier transform:
\begin{equation}
    \mathcal{F}\left(\mathbb{N}_{\noMTFradcpestimate}\right)(\freq)=\left(\mathcal{F}\left(\mathbb{N}_{\hat{\scrad}'_c}\right)G\right)(\freq)
    \label{eq:MTF_15}\,.
\end{equation}
Therefore, the PSD of the noise components of $\noMTFradcpestimate$ are:
\begin{equation}
    \begin{split} 
  N_{\noMTFradcpestimate}(\freq)=\left| \mathcal{F}\left(\mathbb{N}_{\noMTFradcpestimate}\right)(\freq)\right|^2 = \left|\mathcal{F}\left(\mathbb{N}_{\hat{\scrad}'_c}\right)(\freq)G(\freq)\right|^2\\
= \left|\mathcal{F}\left(\mathbb{N} _{\hat{\scrad}'_c}\right)(\freq)\right|^2\left|G(\freq)\right|^2 = N_{\hat{\scrad}'_{c}}(\freq)\left|G\left(\freq\right)\right|^{2}
\end{split}
    \label{eq:MTF_16}.
\end{equation}
By the Wiener–Khinchin theorem, the autocorrelation function $R\left(\tau_{x}, \tau_{y}\right)$ ($\tau_x$,$\tau_y$ are spatial shifts) of a wide-sense stationary random process (assumption (3)) and its PSD $N\left(\freq\right)$ form a Fourier transform pair: 
\begin{equation}
R\left(\tau_{x}, \tau_{y}\right)=\mathcal{F}^{-1}\left(N\left(\freq\right)\right)
\label{eq:MTF_17}.
\end{equation}
Since the noise $\mathbb{N}$ is a zero-mean stochastic process, we have:
\begin{equation}
    \sigma^{2}_{\noMTFradcestimate}(\pix) \approx  R_{\noMTFradcpestimate}(0,0)= \iint N_{\noMTFradcpestimate}(\freq)\, d\freq
\label{eq:MTF_18},
\end{equation}
where second equals sign follows the definition of the inverse Fourier transform. Under the $\hat{\scrad}_{c}$ white noise assumption:
\begin{equation}
    N_{\hat{\scrad}'_{c}}(\freq)\approx\sigma^{2}_{\hat{\scrad}_{c}}(\pix)
    \label{MTF_19},
\end{equation}
considering~\eqref{MTF_16} and~\eqref{MTF_18}, we have:
\begin{equation}
\sigma^{2}_{\noMTFradcestimate}(\pix)\approx\iint N_{\noMTFradcpestimate}(\freq)\, d\freq\approx\sigma^{2}_{\hat{\scrad}'_{c}}(\pix) \iint\left|G\left(\freq\right)\right|^{2} d\freq
\label{eq:MTF_20}.
\end{equation}
The $\approx$ is used because the derivation relies on the three aforementioned assumptions. The accuracy is validated via Monte Carlo simulations in Section~M\ref{main-sec:MC_Valid} in main text.

\section{Color Correction for 4+ primaries}
\label{sec:uncertainty_color_correction}
We derived the color correction method for three-primary displays (RGB subpixels) in the main text. However, some advanced displays aiming for HDR and wide color gamut utilize more than three primaries—for example, OLED displays include an additional white subpixel. This section presents the derivation of the mean and variance for color correction in displays with four or more subpixels.

The radiance $\GHestimate$ recorded by the camera for a single display pixel is:
\begin{equation}
\GHestimate = \sum _\dissubp P_\dissubp\int _{\lambda}E_\dissubp(\lambda)C_c(\lambda)d\lambda
    \label{eq:CC_1},
\end{equation}
where $\lambda$ is the wavelength, $E_\dissubp(\lambda)$ is the spectral power distribution (SPD) of subpixel $\dissubp \in \{\mathsf{R}, \mathsf{G}, \mathsf{B}\}$ at unit intensity, $P_\dissubp$ is the linearized subpixel value, and $C_c(\lambda)$ denotes the spectral sensitivity of camera channel $c \in \{\mathsf{r}, \mathsf{g}, \mathsf{b}\}$. 

Similarly, the trichromatic CIE XYZ value (related to cone responses) can be expressed as:
\begin{equation}
    \ccresult  = \sum _\dissubp P_\dissubp \int_{\lambda }E_\dissubp(\lambda)\bar{\mathbf{S} } (\lambda )d\lambda
    \label{eq:color_xyz}, 
\end{equation}
where $\ccresult=[X,Y,Z]'$ and $\bar{\mathbf{S} } (\lambda ) = [\bar{x}(\lambda ),\bar{y}(\lambda ),\bar{z}(\lambda )]^\top$ are the CIE 1931 standard observer color matching functions (CMFs).

For a display with three color subpixels, there exists a unique transformation matrix $\ccmatrix \in\mathbb{R}^{3 \times 3}$~\footnote{Note that color correction is performed on a per-captured-image-pixel basis.} such that:
\begin{equation}
    \ccestimate(\pixscreen)=\ccmatrix\GHestimateD(\pixscreen)
    \label{eq:color_correction}.
\end{equation}

However, advanced OLED displays employ four subpixels (R, G, B, W) with complex driving mechanisms~\cite{ashraf2024color}, allowing only an approximate transformation matrix to be derived. To improve color correction accuracy, we employed the root polynomial regression method~\cite{finlayson2015color}, which extends $\GHD$ to 
\begin{equation}
\GHDroot(\pixscreen)=\left[\GHD_{\mathsf{r}},\GHD_{\mathsf{g}},\GHD_{\mathsf{b}},\sqrt{\GHD_{\mathsf{r}}\GHD_{\mathsf{g}}},\sqrt{\GHD_{\mathsf{r}}\GHD_{\mathsf{b}}},\sqrt{\GHD_{\mathsf{g}}\GHD_{\mathsf{b}}}\right]^\top(\pixscreen)
\label{eq:CC_3}, 
\end{equation}
and fitted a transformation matrix $\ccmatrixroot \in \mathbb{R}^{3 \times 6}$ such that:
\begin{equation}
\ccestimate(\pixscreen)=\ccmatrixroot\,\GHDroot(\pixscreen)\,, \quad \mu_{\ccestimate}(\pixscreen)=\ccmatrixroot\,\mu_{\GHestimateD}(\pixscreen)
\label{eq:CC_4}.
\end{equation}

To estimate uncertainty, we need to model covariance as the color channels are no longer independent. For the display with more than three primaries, we have used root polynomial regression, and thus apply a first-order Taylor approximation (i.e., Jacobian matrix method) to propagate uncertainty. The Jacobian matrix $J\in \mathbb{R}^{6 \times 3}$ obtained from the root polynomial regression method is:
\begin{equation}
    J=\frac{\partial \GHDroot(\pixscreen)}{\partial\left(\GHD_{\mathsf{r}}(\pixscreen),\GHD_{\mathsf{g}}(\pixscreen),\GHD_{\mathsf{b}}(\pixscreen)\right)} ,\quad J_{ij}=\frac{\partial\left(\GHDroot\right)_{i}(\pixscreen)}{\partial\left(\GHD\right)_{j}(\pixscreen)}
    \label{eq:CC_5}.
\end{equation}
The covariance matrix of $\GHDroot$ and color-corrected result $\ccresult$ are:
\begin{equation}
    \Sigma_{\GHestimateDroot}(\pixscreen)=J\Sigma_{\GHestimateD}(\pixscreen)J^{\top}
    \label{eq:CC_6}
\end{equation}
\begin{equation}
    \Sigma_{\ccestimate}(\pixscreen)=\ccmatrixroot\Sigma_{\GHestimateDroot}(\pixscreen) \ccmatrixroot^{\top}=\ccmatrixroot J\Sigma_{\GHestimateD}(\pixscreen)J^{\top} \ccmatrixroot^{\top}
    \label{eq:CC_6},
\end{equation}
where $\Sigma_{(.)}$ is the covariance matrix.


\bibliographystyle{ACM-Reference-Format}
\bibliography{Camera_Display,refs}